\documentclass[12pt]{iopart}

\usepackage{hyperref}
\usepackage{times}
\expandafter\let\csname equation*\endcsname\relax
\expandafter\let\csname endequation*\endcsname\relax
\usepackage{amsmath}
\usepackage{color}
\usepackage{graphicx}
\usepackage{cite}

\usepackage[section]{placeins}
\usepackage[margin=2cm]{geometry}

\begin{document}
\title{A quantum information processor with trapped ions }
\author{Philipp Schindler$^1$, Daniel Nigg$^1$, Thomas Monz$^1$, Julio T. Barreiro$^{1}$,\\
  Esteban Martinez$^1$, Shannon X. Wang$^2$,  Stephan Quint$^1$, Matthias F. Brandl$^{1}$\\
  Volckmar Nebendahl$^3$, Christian F. Roos$^{4}$, Michael
  Chwalla$^{1,4}$, Markus Hennrich$^{1}$ and Rainer Blatt$^{1,4}$ }

\address{  $^1$Institut f\"ur Experimentalphysik, Universit\"at
 Innsbruck, Technikerstrasse 25, A--6020 Innsbruck,
 Austria\\
 $^2$Massachusetts Institute of Technology, Center for Ultracold Atoms,
Department of Physics, 77 Massachusetts Avenue, Cambridge, MA, 02139, USA \\
 $^3$Institut f\"ur Theoretische Physik, Universit\"at
 Innsbruck, Technikerstrasse 25, A--6020 Innsbruck,
 Austria\\
 $^4$Institut f\"ur Quantenoptik und Quanteninformation
 der \"Osterreichischen Akademie der Wissenschaften,
 Technikerstrasse 21a, A--6020 Innsbruck, Austria}


\begin{abstract}
  Quantum computers hold the promise to solve certain problems
  exponentially faster than their classical counterparts. Trapped
  atomic ions are among the physical systems in which building such a
  computing device seems viable. In this work we present a small-scale
  quantum information processor based on a string of $^{40}$Ca${^+}$
  ions confined in a macroscopic linear Paul trap. We review our set
  of operations which includes non-coherent operations allowing us to
  realize arbitrary Markovian processes. In order to build a larger
  quantum information processor it is mandatory to reduce the error
  rate of the available operations which is only possible if the
  physics of the noise processes is well understood. We identify the
  dominant noise sources in our system and discuss their effects on
  different algorithms. Finally we demonstrate how our entire set of
  operations can be used to facilitate the implementation of
  algorithms by examples of the quantum Fourier transform and the
  quantum order finding algorithm.
\end{abstract}
\tableofcontents
\clearpage
\section{Tools for quantum information processing in ion traps}
\label{sec:tools-quant-inform}
\subsection{Quantum information processing in ion traps}
A quantum computer (QC) hold the promise to solve certain problems
exponentially faster than any classical computer. Its development was
boosted by the discovery of Shor's algorithm to factorize large
numbers and the insight that quantum error correction allows arbitrary
long algorithms even in a noisy
environment~\cite{shor_good,shor-error-correction,Shor1994,steane96}. These
findings initiated major experimental efforts to realize such a
quantum computer in different physical
systems\cite{Blatt2008Entangled,Clarke2008Superconducting,Bloch2008Quantum}. One
of the most promising approaches utilizes single ionized atoms
confined in Paul traps. Here, the internal state of each ion
represents the smallest unit of quantum information (a
qubit). Multiple qubit registers are realized by a linear ion string
and the interaction between different ions along the string is
mediated by the Coulomb
interaction~\cite{Zoller1995,Wineland1995Experimental,hcn_thesis}. In
this work we present a review of a small scale quantum information
processor based on a macroscopic linear Paul
trap~\cite{fsk-decoherence}.  The work is structured as follows: The
first section summarizes the available coherent and non-coherent
operations and in section \ref{sec:experimental-setup} the
experimental setup is reviewed. In section \ref{sec:error-sources} the
noise sources are characterized, and finally, in section
\ref{sec:using-toolb-quant} we discuss examples of implemented
algorithms that use the full set of operations.

\subsection{The qubit - $^{40}$Ca$^+$ }
\begin{figure}[hh]
  \centering
  \includegraphics[width=15cm]{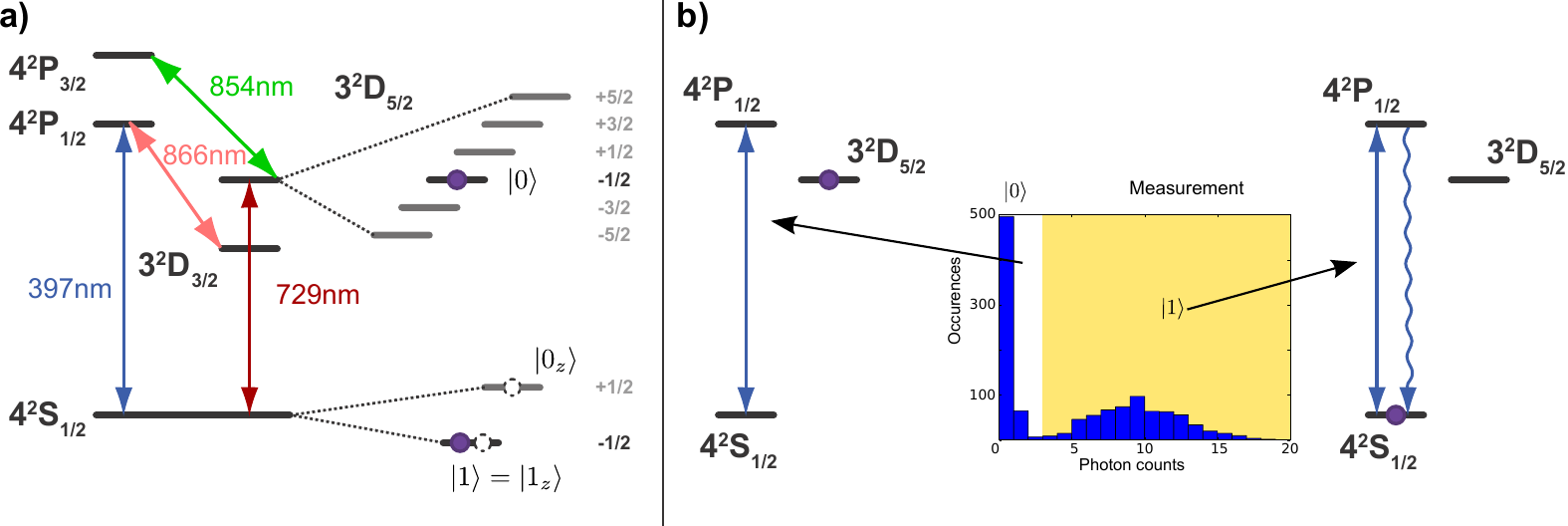}
  \caption{(a) Level scheme of $^{40}$Ca$^+$. Solid circles indicate
    the usual optical qubit ($4S_{1/2}(m_j=-1/2) = |1\rangle$ and
    $3D_{5/2} (m_j=-1/2)= |0\rangle$). Open circles indicate the
    ground state qubit which is not subject to spontaneous decay
    ($4S_{1/2}(m_j=-1/2) = |1\rangle_Z$ and $4S_{1/2}(m_j=+1/2)=
    |0\rangle_Z$). (b) Schematic representation of electron shelving
    detection. The histogram shows the detected photon counts from
    projections onto both states during the detection interval. It can
    be seen that it is possible to distinguish the two different
    outcomes. The highlighted area illustrates the threshold whether
    the state is detected as $|0\rangle$ or $|1\rangle$.}
  \label{fig:LevelScheme}
\end{figure}

A crucial choice for any QC implementation is the encoding of a qubit
in a physical system.  In ion trap based QCs, two distinct types of
qubit implementations have been explored: (i) ground-state qubits
where the information is encoded in two hyperfine or Zeeman sublevels
of the ground state~\cite{Wineland1995Experimental}, and (ii) optical
qubits where the information is encoded in the ground state and an
optically accessible metastable excited
state~\cite{fsk-decoherence}. The two types of qubits require distinct
experimental techniques where, in particular, ground-state qubits are
manipulated with either two-photon Raman transitions or by direct
microwave excitation~\cite{Wineland1995Experimental}. In contrast,
operations on optical qubits are performed via a resonant light field
provided by a laser~\cite{roos-decoherence,fsk-decoherence}.
Measuring the state of the qubits in a register is usually performed
by the electron shelving method using an auxiliary short-lived state
for both qubit types~\cite{Wineland1995Experimental}. In the presented
setup we use $^{40}$Ca$^+$ ions, which contain both, an optical qubit
for state manipulation and a ground-state qubit for a quantum
memory. Figure~\ref{fig:LevelScheme}a) shows a reduced level scheme of
$^{40}$Ca$^+$ including all relevant energy levels.

Our standard qubit is encoded in the $4S_{1/2}$ ground state and the
$3D_{5/2}$ metastable state, where the natural lifetime of the
$3D_{5/2}$ state ($\tau_1 = 1.1s$) provides an upper limit to the
storage time of the quantum information.  The $4S_{1/2}$ state
consists of two Zeeman sublevels ($m=\pm 1/2$) whereas the $3D_{5/2}$
state has six sublevels ($m=\pm 1/2, \pm 3/2, \pm 5/2$). This leads to
ten allowed optical transitions given the constraint that only $\Delta
m=0,1,2$ are possible on a quadrupole transition. The coupling
strength on the different transitions can be adjusted by varying the
polarization of the light beam and its angle of incidence with respect
to the quantization axis set by the direction of the applied magnetic
field.  Usually we choose the $4S_{1/2}(m_j=-1/2)=|S\rangle =
|1\rangle$ and the $3D_{5/2} (m_j=-1/2)=|D\rangle= |0\rangle$ as the
computational basis states because the transition connecting them is
the least sensitive to fluctuations in the magnetic field. Furthermore
it is possible to store quantum information in the two Zeeman
substates of the $4S_{1/2}$ ground-state which are not subject to
spontaneous decay: $4S_{1/2}(m_j=-1/2) = |1\rangle_Z$ and
$4S_{1/2}(m_j=+1/2)= |0\rangle_Z$.

The projective measurement of the qubit in the computational basis is
performed via excitation of the $4S_{1/2} \leftrightarrow 4P_{1/2}$
transition at a wavelength of 397nm. If the qubit is in a
superposition of the qubit states, shining in a near resonant laser at
the detection transition projects the ion's state either in the
$4S_{1/2}$ or the $3D_{5/2}$ state. If the ion is projected into the
$4S_{1/2}$ state, a closed cycle transition is possible and the ion
will fluoresce as sketched in figure~\ref{fig:LevelScheme}(b). It is
however still possible that the decay from $4P_{1/2}$ leads to
population being trapped in the $3D_{3/2}$ state that needs to be
pumped back to the $4P_{1/2}$ with light at
866nm~\ref{fig:LevelScheme}(b). Fluorescence is then collected with
high numerical aperture optics and single-photon counting devices as
described in section~\ref{sec:experimental-setup}. If the ion is
projected into the $3D_{5/2}$ state though, it does not interact with
the light field and no photons are scattered.  Thus the absence or
presence of scattered photons can be interpreted as the two possible
measurement outcomes which can be clearly distinguished as shown in
the histogram in figure~\ref{fig:LevelScheme}b).  In order to measure
the probability $p_{|1\rangle}$ to find the qubit in $4S_{1/2}$, this
measurement needs to be performed on multiple copies of the same
state.  In ion-trap QCs these multiple copies are produced by
repeating the experimental procedure $N$ times yielding the
probability $p_{|1\rangle} = n(|1\rangle)/N$ where $n(|1\rangle)$ is
the number of bright outcomes.  This procedure has a statistical
uncertainty given by the projection noise $\Delta p_{|1\rangle} =
\sqrt{p_{|1\rangle} (1-p_{|1\rangle}) / N} $~\cite{Itano1993Quantum}.
Depending on the required precision, the sequence is therefore
executed between 50 and 5000 times.

\begin{figure}[hh]
  \centering
  \includegraphics[width=10.5cm]{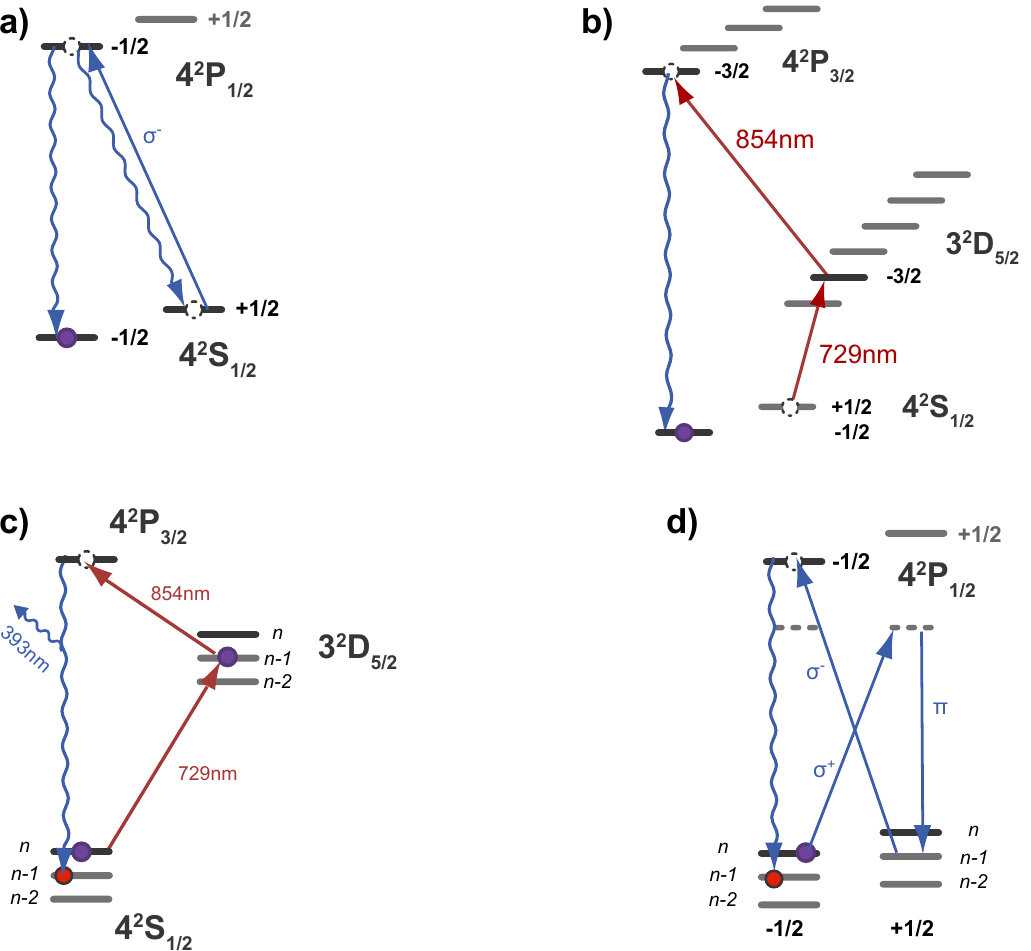}     
  \caption{Schematic view of optical pumping which is (a) polarization
    selective and (b) frequency selective (c) Sideband cooling on the
    qubit transition. The light resonant with the $3D_{5/2}
    \rightarrow 4P_{3/2}$ transition is used to tune the effective
    linewidth of the excited state leading to an adiabatic elimination
    of the $3D_{5/2}$ state. (d) Scheme for sideband cooling utilizing
    a Raman transition. Here, the $\sigma^-$ light performs optical
    pumping which corresponds to the spontaneous decay on the optical
    transition.}
  \label{fig:OpticalPumpingSidebandCool}
\end{figure}

Preparing the qubit register in a well defined state is a crucial
prerequisite of any quantum computer. In our system this means (i)
preparing the qubit in one of the two Zeeman levels of the ground
state and (ii) cooling the motional state of the ion string in the
trap to the ground state. The well established technique of optical
pumping is used to prepare each ion in the $m_j=-1/2$ state of the
$4S_{1/2}$ state~\cite{fsk-decoherence}. In our setup two distinct
methods for optical pumping are available: (i) Polarization dependent
optical pumping by a circularly polarized laser beam resonant on the
$4S_{1/2} \leftrightarrow 4P_{1/2}$ transition as shown in
figure~\ref{fig:OpticalPumpingSidebandCool}a) and (ii) frequency
selective optical pumping via the Zeeman substructure of the
$3D_{5/2}$ state as depicted in
figure~\ref{fig:OpticalPumpingSidebandCool}b). Here, the transfer on
the qubit transition at 729~nm is frequency selective. Selection rules
ensure that depletion of the $3D_ {5/2}(m_j=-3/2)$ level via the
$4P_ {3/2}$ effectively pumps the population into the
$4S_{1/2}(m_j=-1/2)$ state.
The second part of the initialization procedure prepares the ion
string in the motional ground state which requires multiple
laser-cooling techniques.  We use a two-step process where the first
step consists of Doppler cooling on the $4S_{1/2} \leftrightarrow
4P_{1/2}$ transition leading to a mean phonon number of $\langle n
\rangle \approx 10$. The motional ground state is subsequently reached
with sideband cooling techniques~\cite{Eschner2003Laser}. In our
system, the necessary two-level system can be either realized on the
narrow qubit transition~\cite{Rohde2001Sympathetic} or as a Raman
process between the two ground states via the $4P_{1/2}$
level~\cite{Wineland1995Experimental,fsk-decoherence}.
A crucial parameter, determining the cooling rate, is the linewidth of
the actual cooling transition~\cite{Eschner2003Laser}. When cooling on
the long-lived optical transition, the excited state lifetime needs to
be artificially shortened in order to adjust the effective linewidth
of the transition. This is realized by repumping population from the
$3D_{5/2}$ state to the $4S_{1/2}$ state via the $4P_{3/2}$ level with
light at 854nm, as outlined in
figure~\ref{fig:OpticalPumpingSidebandCool}c)~\cite{Eschner2003Laser}. The
procedure using the Raman transition is outlined in
figure~\ref{fig:OpticalPumpingSidebandCool}d). Here, the spontaneous
decay is replaced by optical pumping as used for state
preparation~\cite{Wineland1995Experimental,Marzoli1994Laser}. In
principle, this cooling technique allows for faster cooling rates as
the coupling strength to the motional mode, described by the
Lamb-Dicke parameter, increases for smaller wavelengths. More
importantly, it has the advantage of being applicable within a quantum
algorithm without disturbing the quantum state of idling qubits when
the population of the $4S_{1/2}(m_j=-1/2)=|0\rangle$ state is
transferred to a Zeeman substate of the excited state that is outside
the computational basis, for example $3D_{5/2}
(m_j=-5/2)=|D'\rangle$~\cite{mark_tele}.
\subsection{The universal set of gates}

\begin{figure}[hh]
  \centering
  \includegraphics[width=14.5cm]{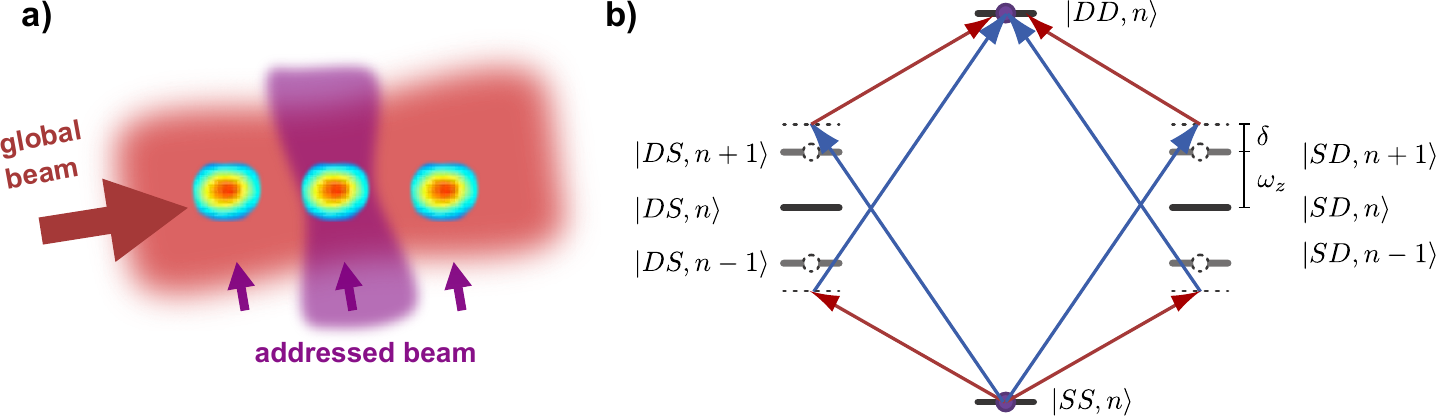}    
  \caption{a) Schematic view of the laser beam geometry for qubit
    manipulation.  b) Schematic level scheme of a M{\o}lmer
    S{\o}rensen type interaction. The bichromatic light field couples
    the states $|SS,n\rangle$ with $|DD,n\rangle$ via the intermediate
    states $|SD,n \pm 1\rangle$ and $|DS,n \pm 1\rangle$ with a
    detuning $\delta$. }
  \label{fig:MSlevel}
\end{figure}

With a universal set of gates at hand, every unitary operation acting
on a quantum register can be implemented~\cite{nielsen_chuang}. The
most prominent example for such a set consists of arbitrary
single-qubit operations and the controlled NOT (CNOT) operation.
However, depending on the actual physical system, the CNOT operation
may be unfavorable to implement and thus it may be preferable to
choose a different set of gates. In current ion trap systems,
entangling operations based on the ideas of M{\o}lmer and S{\o}rensen
have achieved the highest
fidelities~\cite{ms_jan,molmer_sorensen,molmer}. These gates, in
conjunction with single-qubit operations, form a universal set of
gates. In order to implement all necessary operations, we use a wide
laser beam to illuminate globally the entire register uniformly and a
second, tightly focused, steerable laser beam to address each
ion. Interferometric stability between the two beams would be
required, if arbitrary single-qubit operations were performed with
this addressed beam in addition to the global operations. To
circumvent this demanding requirement, the addressed beam is only used
for inducing phase shifts caused by the AC-Stark effect. Using such an
off-resonant light has the advantage that the phase of the light field
does not affect the operations and thus no interferometric stability
is needed. The orientation of the two required laser beams is shown in
figure~\ref{fig:MSlevel}a).

Applying an off-resonant laser light with Rabi frequency $\Omega$ and
detuning $\delta$ onto a the $j$-th ion modifies its qubit transition
frequency by an AC-Stark shift of $\delta_{AC}=-\frac{\Omega^2}{2
  \Delta}$.  This energy shift causes rotations around the Z axis of
the Bloch sphere and the corresponding operations on ion $j$ can be
expressed as
\[ S_z^{(j)}(\theta) = e^{-i \theta \sigma_z^{(j)} /2 } \] where the
rotation angle $\theta = \delta_{AC} t$ is determined by the AC-Stark
shift and the pulse duration.  As the $^{40}$Ca$^+$ ion is not a
two-level system, the effective frequency shift originates from
AC-Stark shifts on multiple transitions. We choose the laser frequency
detuning from any $4S_{1/2} \leftrightarrow 3D_{5/2}$ transition to be
20MHz. There, the dominating part of the AC-Stark shift originates
from coupling the far off-resonant transitions from $4S_{1/2}$ to
$4P_{1/2}$ and $4P_{3/2}$ as well as from $3D_{5/2}$ to
$4P_{3/2}$~\cite{haeffner_2003_stark}.

The second type of non-entangling operations are collective resonant
operations using the global beam. They are described by
$$ R_{\phi}(\theta)  = e^{-i \theta  S_\phi / 2}$$ 
where $S_\phi = \sum_{i=0}^N (\sigma_x^{(i)} \cos \phi +
\sigma_y^{(i)} \sin \phi)$ is the sum over all single-qubit Pauli
matrices $\sigma_{x,y}^{(i)}$ acting on qubit $i$. The rotation axis
on the Bloch sphere $\phi$ is determined by the phase of the light
field and the rotation angle $\theta = t \, \Omega$ is fixed by the
pulse duration $t$ and the Rabi frequency $\Omega$. Together with the
single-qubit operations described above this set allows us to
implement arbitrary non-entangling operations on the entire register.

The entangling MS gate operation completes the universal set of
operations. The ideal action of the gate on an N-qubit register is
described by
$$ MS_\phi(\theta) = e^{-i \theta S_\phi^2/4 } \; .$$
For any even number of qubits the operation $MS_\phi(\pi/2)$ maps the
ground state $|00..0\rangle$ directly onto the maximally entangled GHZ
state $1/\sqrt{2} (|00..0\rangle - i \, e^{i N \phi} |11..1\rangle )$. For an
odd number of ions the produced state is still a maximally GHZ-class
entangled state which can be transferred to a GHZ state by an
additional collective local operation $R_{\phi}(\pi/2)$. 

Implementing the MS gate requires the application of a bichromatic
light field $E(t) = E_+(t) + E_-(t) $ with constituents $E_\pm = E_0
\cos ((\omega_0 \pm (w_z + \delta)) t)$ where $\omega_0$ is the qubit
transition frequency, $\omega_z$ denotes the frequency of the motional
mode and $\delta$ is an additional detuning. The level scheme of the
MS operation acting on a two-ion register is shown in
figure~\ref{fig:MSlevel}b). M{\o}lmer and S{\o}rensen showed that if
the detuning from the sideband $\delta$ equals the coupling strength
on the sideband $\eta \Omega$ the operation $MS(\pi / 2)$ is performed
when the light field is applied for a duration $t=2 \pi /
\delta$. 

However, implementing MS operations with rotation angles $\pi/2$ is
not sufficient for universal quantum computation. Arbitrary rotation
angles $\theta$ can be implemented with the same detuning $\delta$ by
adjusting the Rabi frequency on the motional sideband to $\eta \Omega=
\delta \, \sqrt{ \theta  \, / (\pi/2)}$.  
Due to this fixed relation between the
rotation angle and the detuning, the gate operation needs to be
optimized for each value of $\theta$. In practice this optimization is
a time-consuming task and thus the gate is optimized only for the
smallest occurring angle in the desired algorithm. Gate operations
with larger rotation angles are realized by a concatenation of
multiple instances of the already optimized operation.

If the physical system consisted of a two-level atom coupled to a
harmonic oscillator the AC-Stark introduced by one off-resonant light
field would be perfectly compensated by its counterpart in the
bichromatic field. However, $^{40}$Ca$^+$ shows a rich level structure
where due to the additional Zeeman levels and coupling to the other
$4P$ states an additional AC-Stark shift is
introduced~\cite{haeffner_2003_stark}. This shift changes the
transition frequency between the two qubit states which has the effect
that the detuning from the sideband transition $\delta$ is not equal
for both constituents of the bichromatic light field. This would
degrade the quality of the operation drastically and thus the shift
has to be compensated which can be achieved by two distinct
techniques~\cite{thermal_ms}: (i) The center frequency of the
bichromatic light field can be shifted or (ii) the light intensity of
the two constituents can be unbalanced to induce a Stark shift on the
carrier transition which compensates the unwanted Stark shift.
Depending on the application, one compensation method is preferable
over the other. Method (i) makes it easier to optimize the physical
parameters to achieve very high gate fidelities but leads to an
additional global rotation around $\sigma_z$ which is tedious to
measure and compensate for in a complex algorithm. This can be avoided
by method (ii) but the compensation is not independent of the motional
state leading to a slightly worse
performance~\cite{thermal_ms}. Therefore, we generally choose method
(i) if the goal is to solely generate a GHZ state whereas method (ii)
is favorable if the gate is part of a complex algorithm.

In general an algorithm requires operations with positive and negative
values of the rotation angles for the available operations. For the
resonant $R_\phi(\theta)$ operation both signs of $\theta$ can be
realized by changing the phase of the light field since $e^{-i
  (-\theta) S_{\phi}} = e^{-i\theta S_{(\pi+\phi)}}$ which is not
possible for MS operations as $S_\phi^2 = S_{\phi+\pi}^2$. The sign of
the rotation of the MS operation angle can only be adjusted by
choosing the sign of the detuning
$\delta$~\cite{Muller2011Simulating}.  However, performing MS
operations with positive and negative detunings results in a more
complex setup for generating the required RF signals and also a
considerable overhead in calibrating the operation. Therefore it can
be favorable to implement negative $\theta$ by performing
$MS_{\phi}(\pi-|\theta |)$ which works for any odd number of ions
whereas for an even number of ions, an additional $R_\phi(\pi)$
operation is required~\cite{Muller2011Simulating}. With this approach
the quality of operations with negative rotation angles is reduced but
the experimental overhead is avoided.


\subsection{Optimized sequences of operations}

Typically, quantum algorithms are formulated as quantum circuits where
the algorithm is build up from the standard set of operations
containing single qubit operations and CNOT gates. Implementing such
an algorithm is straightforward if the implementation can perform
these standard gate operations efficiently.  Our set of gates is
universal and thus it is possible to build up single qubit and CNOT
operations from these gates. However, it might be favorable to
decompose the desired algorithm directly into gates from our
implementable set as the required sequence of operations might require
less resources. This becomes evident when one investigates the
 operations necessary to generate a four-qubit GHZ state. Here, a single
MS gate is able to replace four CNOT gates.

The problem of breaking down an algorithm into an optimized sequence
of given gate operations was first solved by the NMR quantum computing
community. There, a numerical optimal control algorithm was employed
to find the sequence of gate operations that is expected to yield the
lowest error rate for a given unitary operation~\cite{Khaneja2005}.
This algorithm optimizes the coupling strength of the individual parts
of the Hamiltonian towards the desired sequence. Unfortunately the NMR
algorithm is not directly applicable to our ion trap system as the set
of operations differ. In an NMR system the interactions are present at
all times, only their respective strengths can be controlled. This
allows for an efficient optimization as there is no time order of the
individual operations. This is not true for current ion trap quantum
computers where only a single operation is applied at a time which
makes it necessary to optimize the order of the operations within the
sequence in addition to the rotation angles. Furthermore, the same
type of operation might appear several times at different
positions. Thus we modified the algorithm so that it starts from a
long random initial sequence and optimizes the rotation angles of the
operation. This optimization converges towards the desired algorithm,
if the required sequence is a subset of this random initial
sequence. The key idea of our modification is that rotation angles of
operations that are included in the random initial sequence but are
not required for the final sequence shrinks during the
optimization. If the rotation angle of an operation shrinks below a
threshold value, the operation is removed from the sequence as it is
superfluous.  If the algorithm fails to find a matching sequence,
further random operations are inserted into the sequence. A more
detailed treatment on the algorithm is given in
reference~\cite{Volckmar}. In general this optimization method is not
scalable as the search space increases exponentially with the number
of qubits but it is possible to build up an algorithm from optimized
gate primitives acting on a few qubits.

Even for complex algorithms on a few qubits, the sequence generated
with this optimization method might include too many operations to
yield acceptable fidelities when implemented. Then it can be
advantageous to split the algorithm in parts that act only on a subset
of the register and generate optimized decompositions for these
parts. For this task, the physical interactions need to be altered so
that they only affect the relevant subset. Multiple techniques for
achieving this in ion traps have been proposed, where the best known
techniques rely on physically moving and splitting the ion-chains in a
complex and miniaturized ion trap~\cite{wineland_scaling}.  Our
approach to this problem is to decouple them spectroscopically by
transferring the information of the idling ions into a subspace that
does not couple to the resonant laser light.  Candidates for such
decoupled subspaces are either (i) $4S_{1/2} (m_j=+1/2)$ with
$3D_{5/2} (m_j=+1/2)$ or alternatively (ii) $3D_{5/2}
(m_j=-5/2)=|D'\rangle$ with $3D_{5/2} (m_j=-3/2)=|D''\rangle$.  The
decoupling technique (ii) is sketched in
figure~\ref{fig:SchematicReset}a).  The only remaining action of the
manipulation laser on the decoupled qubits is then an AC-Stark shift
that acts as a deterministic rotation around the Z-axis. This rotation
can be measured and subsequently be compensated for by controlling the
phase of the transfer light.  When qubits in the set $U$ are
decoupled, the action of the operations can then be described by
$(\prod_{j \in U} {1}_j) \otimes U$ where the operation $U$ is the
implemented interaction on the desired subspace.  Note that the
parameters of the MS operations do not change when the number of
decoupled qubits is altered thus the gate does not need to be
re-optimized.
\subsection{Tools beyond coherent operations}

In general, any quantum computer requires non-reversible and therefore
also non-coherent operations for state initialization and
measurements~\cite{nielsen_chuang}.  For example, quantum error
correction protocols rely on controlled non-coherent operations within
an algorithm to remove information on the error from the system
similar to state initialization. Furthermore, the robustness of a
quantum state against noise can be analyzed by exposing it to a well
defined amount of phase or amplitude damping~\cite{bound}.
Surprisingly, it has been shown theoretically that non-coherent
operations can serve as a resource for quantum
information~\cite{Verstraete2009Quantum,Diehl2008Quantum,Muller2011Simulating}.
Naturally, these ideas can only be implemented if controlled
non-coherent operations are available in the system.  Mathematically,
these non-reversible operations are described by a trace-preserving
completely positive map $\mathcal{E}(\rho)$ acting on a density matrix
rather than unitary operations acting on pure states. The action of
such a map is described by $ \mathcal{E}(\rho) = \sum_k E_k^\dagger
\rho E_k$ with Kraus operators $E_k$ fulfilling $\sum_k E_k^\dagger
E_k = 1$~\cite{nielsen_chuang}.

\begin{figure}[hh]
  \centering
  \includegraphics[width=16cm]{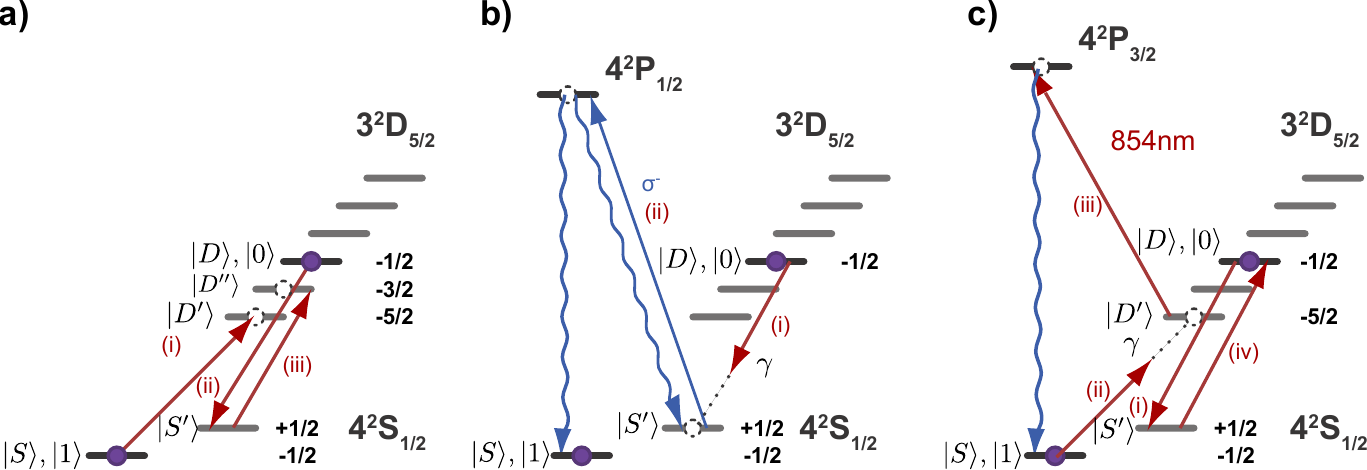}    
  \caption{ a) The process to decouple individual qubits: (i) The
    population from $|S\rangle$ is transferred to $|D'\rangle$. (ii)
    The population from $|D\rangle$ is transferred to $|S'\rangle$ and
    subsequently to (iii) $|D''\rangle$. b) Implementing controlled
    amplitude damping using the 397nm $\sigma$ beam. (i) Transferring
    the population from $|D\rangle$ to $|S'\rangle$. (ii) Optical
    pumping of $|S'\rangle$ using light at 397nm. c) Controlled phase
    damping with strength $\gamma$ utilizing light at 854nm. (i)
    Population from $|D\rangle$ is hidden in the $|S'\rangle$
    state. (ii) The population from $|S\rangle$ is partially brought
    to $|D'\rangle$ and (iii) shining in light at 854 nm depletes the
    $3D_{5/2}$ via $4P_{3/2}$ and finally (iv) the population is
    brought from $|S'\rangle$ back to $|D\rangle$.  }
  \label{fig:SchematicReset}
\end{figure}

In our system two different variations of these controlled dissipative
processes are available~\cite{Nigg2013Experimental}: The archetype of
a controlled non-coherent optical process is optical pumping. We can
perform optical pumping on individual qubits inside the register with
the following sequence as shown in figure~\ref{fig:SchematicReset}c):
(i) Partially transfer the population from $|D\rangle$ to $|S'\rangle$
with probability $\gamma$, and (ii) optical pumping from $|S'\rangle$
to $|S\rangle$ analogous to the qubit initialization. The partial
population transfer is performed by a coherent rotation with an angle
$\theta$ on the transition $4S_{1/2}(m_j=+1/2) \leftrightarrow
3D_{5/2}(m_j=-1/2) $ which leads to $\gamma = \sin^2 (\theta)$. This
reset process can be described as controlled amplitude damping on an
individual qubit where the map affecting the qubit is shown in
table~\ref{tab:SetOfOperations}. Note that the information in the
qubit states is not affected as the optical pumping light couples to
neither of the original qubit states. For a full population transfer
($\gamma=1$) the procedure acts as a deterministic reinitialization of
an individual qubit inside a register as required for repetitive
quantum error correction~\cite{Schindler2011Experimental}.

Furthermore an alternative implementation of optical pumping can be
used to generate controlled phase damping. This process preserves the
populations in the respective qubit states but destroys the coherences
between them with probability $\gamma$: (i) The information residing
in state $|D\rangle$ of all qubits is protected by transferring it to
the $|S'\rangle= 4S_{1/2} (m_j=+1/2)$ state before the reset
step. (ii) On the qubit to be damped, the population from $|S\rangle$
is partially transferred into the $|D'\rangle= 3D_{5/2} (m_j=-5/2)$
state with probability $\gamma$. Here, the partial population transfer
is performed by a coherent rotation on the transition
$4S_{1/2}(m_j=-1/2) \leftrightarrow 3D_{5/2}(m_j=-5/2) $ (iii) Shining
light resonant with the $3D_{5/2} \leftrightarrow 4P_{3/2}$ transition
at 854~nm onto the ions depletes this level to $|S\rangle$. (iv)
Transferring $|S'\rangle$ back to $|D\rangle$ restores the initial
populations, the coherence of the qubit has been destroyed with
probability $\gamma$.  The schematic of this process is shown in
figure \ref{fig:SchematicReset}b) and the resulting map can be found
in table~\ref{tab:SetOfOperations}.

Our system furthermore allows the measurement of a single qubit
without affecting the other qubits in the same ion string. For this, all
spectator ions need to be decoupled from the detection light. This is
realized by transferring the population from the $|S\rangle$ state to
the $|D'\rangle = 3D_{5/2} (m_j=-5/2)$ state. Applying light on the
detection transition measures the state of the ion of interest while
preserving the quantum information encoded in the hidden qubits. This
information can be used to perform conditional quantum operations as
needed for teleportation experiments~\cite{mark_tele} or quantum
non-demolition measurements~\cite{Barreiro2011Opensystem}.

It should be noted, that the operations forming our implementable set
of gates shown in table \ref{tab:SetOfOperations} allow the
realization of any completely positive map, which corresponds to a
Markovian
process~\cite{Baggio2012Quantum,Lloyd2001Engineering,Barreiro2011Opensystem}.
The quality of the operations is affected by multiple physical
parameters that are discussed in more detail in
section~\ref{sec:error-sources}. In order to faithfully estimate the
resulting fidelity of an implemented algorithm, a complete numerical
simulation of the physical system has to be performed.  However, a
crude estimation can be performed assuming a fidelity of $99.5\%$ for
non-entangling operations and $\{98,97,95,93,90\} \%$ for the MS
operations on a string of $\{2,3,4,5,6\}$ ions~\cite{GHZ}. The
fidelity of the entire algorithm is then estimated by simply
multiplying the fidelities of the required operations.

\begin{table}[hh]
  \centering
  \caption{The extended set of operations in our ion trap QC.
    This set of operations allows us to implement any possible Markovian process.}
  \begin{tabular}{c|c|c}
    Name & Addressed/global & Ideal operation \\  \hline
    AC-Stark shift pulses & addressed &  $ S_z^{(i)}(\theta) = e^{-i \theta / 2  \sigma_z^{(i)}}$ \\ \hline
    Collective resonant operations & collective non-entangling &   $S_{\phi}(\theta)  = e^{-i \theta / 2 S_\phi}$ \\ \hline
    M{\o}lmer-S{\o}rensen & collective entangling & $MS_\phi(\theta) = e^{-i \theta/2 S_\phi^2}$\\ \hline
    Phase damping & addressed non-coherent &  $E_0^p = \bigl{|}\begin{smallmatrix} 1 & 0 \\ 0 & \sqrt{1-\gamma} \end{smallmatrix} \bigr{|}$
    $E_1^p = \bigl{|}\begin{smallmatrix} 0 & 0 \\ 0 &  \sqrt{\gamma} \end{smallmatrix} \bigr{|}$ \\ \hline
    Amplitude damping & addressed non-coherent & $E_0^a = \bigl{|}\begin{smallmatrix} 1 & 0 \\ 0 & \sqrt{1-\gamma} \end{smallmatrix} \bigr{|}$
    $E_1^a = \bigl{|}\begin{smallmatrix} 0 & \sqrt{\gamma} \\ 0 & 0 \end{smallmatrix} \bigr{|}$ \\ \hline
    Single-qubit measurement & addressed non-coherent & Projection onto $|0\rangle \langle 0|$ or $|1\rangle \langle 1|$ \\ \hline
  \end{tabular}
  \label{tab:SetOfOperations}
\end{table}

\section{Experimental setup}
\label{sec:experimental-setup}
In this section we give an overview of the experimental setup of our
ion-trap quantum information processor. First, we describe in detail
the ion trap, the optical setup and the laser sources. Then we
concentrate on the experiment control system and techniques to infer
the state of the qubit register.  

\subsection{The linear Paul trap}
\begin{figure}[hh]
  \centering
  \includegraphics[width=15cm]{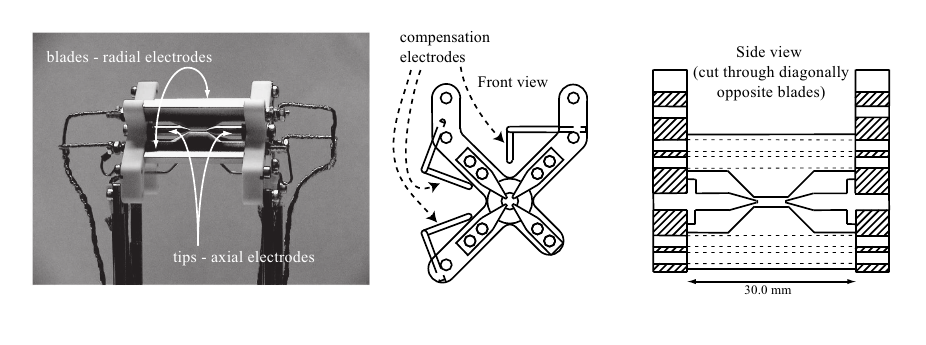}    
  \caption{Schematic drawing of the linear Paul trap used in our
    experiment. The distance between the endcaps is 5mm whereas the
    distance between the radio-frequency blades is 1.6mm.}
  \label{fig:Trap}
\end{figure}

The trap in our experimental system is a macroscopic linear Paul trap
with dimensions as shown in
figure~\ref{fig:Trap}\cite{fsk-decoherence}.  The trap is usually
operated at a radial motional frequency of 3MHz and an axial motional
frequency of 1MHz. These trapping parameters are slightly adjusted
with respect to the number of ions in the string to prevent overlap of
the frequencies from different motional modes of all transitions. In
order to minimize magnetic field fluctuations, the apparatus is
enclosed in a magnetic shield (75x75x125~cm) that attenuates the
amplitude of an external magnetic field at frequencies of above 20 Hz
by more than 50dB~\footnote{Imedco, Proj.Nr.: 3310.68}.  The trap
exhibits heating rates of 70ms per phonon at an axial trap frequency
of 1MHz. Micromotion of a single ion can be compensated with the aid
of two compensation electrodes. The remaining micromotion creates
sidebands at the trap frequency which can be observed in an ion
spectrum on the qubit transition. The strength of the excess
micromotion is described by the modulation index~$\beta$ of these
sidebands where in our setup a modulation index of $\beta < 1\%$ is
observed~\cite{Berkeland1998Minimization,Chwalla2009Absolute}.

\subsection{Optical setup}
A quantum information processor with $^{40}$Ca$^+$ requires multiple
laser sources, listed in table~\ref{tab:LaserWavelength}, to prepare,
manipulate and measure the quantum state of the ions. The ions are
generated from a neutral atom beam with a two-step photo-ionization
process requiring laser sources at 422nm and 375nm. Manipulating the
state of the qubits is done with a Titanium-Sapphire laser at 729nm on
the $4S_{1/2} \leftrightarrow 3D_{5/2}$ qubit transition and its setup
as described in reference~\cite{chwalladiss}. Its frequency and
amplitude fluctuations affect crucially the performance of the
coherent operations as will be discussed in
section~\ref{sec:error-sources}. The laser has a linewidth of below
20Hz and the relative intensity fluctuations are in the range of
1.5\%~\cite{chwalladiss}.
\begin{table}[hh]
  \centering
  \begin{tabular}{c|c|c|c}
    Transition & Wavelength & Usage & Linewidth  \\ \hline
    $4S_{1/2} \leftrightarrow 4P_{1/2}$ & 397nm & Doppler cooling, optical pumping and detection & $<$1MHz \\ \hline
    $4S_{1/2} \leftrightarrow 3D_{5/2}$ & 729nm & Sideband cooling and qubit manipulation  & $<$ 20Hz\\ \hline
    $3D_{3/2} \leftrightarrow 4P_{1/2}$ & 866nm & Repumping for detection  & $<$1MHz \\ \hline
    $3D_{5/2} \leftrightarrow 4P_{3/2}$ & 854nm & Quenching for Sideband cooling and qubit reset  & $<$1MHz\\ \hline
    neutral calcium & 422nm & Photoionization first stage  & - \\ \hline
    neutral calcium & 375nm & Photoionization second stage & - \\ \hline
  \end{tabular}
  \caption{Laser wavelengths needed for a Ca$^+$ ion trap experiment. 
    The lasers are stabilized to a reference cavity with the Pound-Drever-Hall locking technique except for
    the photoionization lasers which are not actively stabilized.}
  \label{tab:LaserWavelength}
\end{table}

\begin{figure}[hh]
  \centering
  \includegraphics[width=12cm]{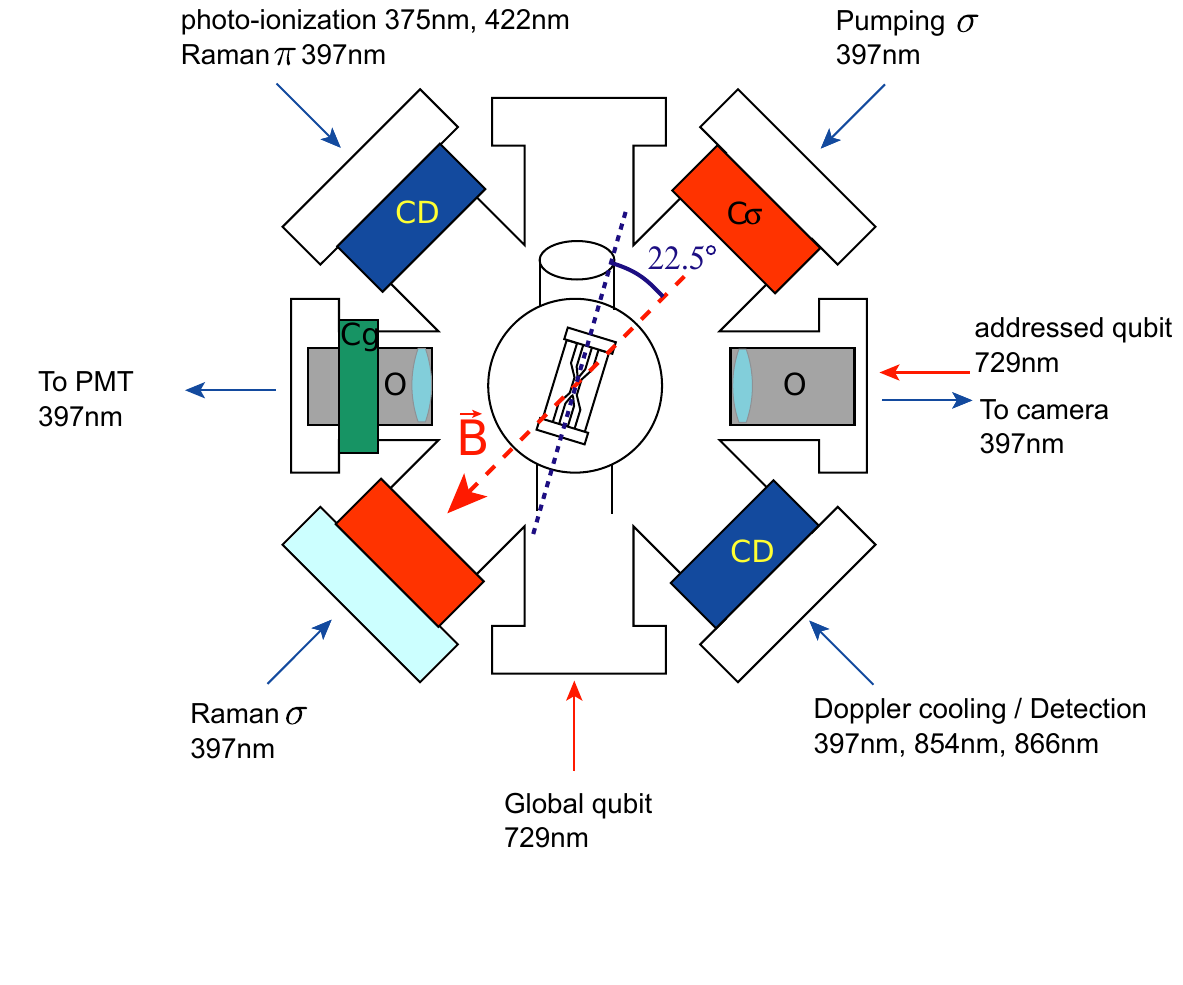}    
  \caption{Overview of the alignment of the various laser beams, the
    coils generating the magnetic field and the trap with respect to
    the vacuum vessel.}
  \label{fig:OpticTrap}
\end{figure}

The vacuum vessel housing the trap and the laser sources sources
reside on different optical tables and thus the light is transferred
between different tables with optical fibers.  The optical access to
the trap itself is constrained by the surrounding octagon vacuum
vessel which is sketched in figure~\ref{fig:OpticTrap} including the
available beams with their respective directions. The 397nm light is
required for multiple tasks and thus multiple beams are required: one
beam for Doppler cooling and detection, another beam for optical
pumping (labeled Pumping $\sigma$), and two beams for Raman
sideband-cooling (labeled Raman $\sigma$, Raman $\pi$).  In
particular, the beams used for optical pumping need to be aligned with
the magnetic field generated by the coils as indicated in
figure~\ref{fig:OpticTrap}.  In practice it is favorable to adjust the
orientation of the magnetic field with respect to the light beam since
the magnetic field can be adjusted without moving any mechanical
part. The beams of the 866nm and 854nm laser are overlapped with the
397nm detection beam in a single-mode photonic crystal fiber.

In order to implement our set of operations, the 729~nm light needs to
be applied to the ions from two different optical ports: (i) the
addressed beam which is a tightly focused beam illuminating only a
single ion and (ii) the global beam which is a wide beam that
illuminates all ions with an approximately homogeneous light
intensity.  The angle between the global beam and the axial trap axis
is $22.5^{\circ}$ which leads to a Lamb-Dicke parameter of
$\eta_{glob}=6 \%$~\cite{Gulde_diss}. The width of the beam is chosen
so that the light intensity shows variations of less than 2\% over the
entire ion string. Considering that the ions are arranged in a linear
crystal, it is advantageous to use an elliptical shape for the global
beam to achieve higher light intensities at the position of the
ions. The elongated axis of the beam has typically a diameter of $100
\mu m$ which is sufficient for ion strings with up to 8 ions. For
larger ion strings, the beam size needs to be enlarged which increases
the required time for performing collective operations.

\begin{figure}[hh]
  \includegraphics[width=17cm]{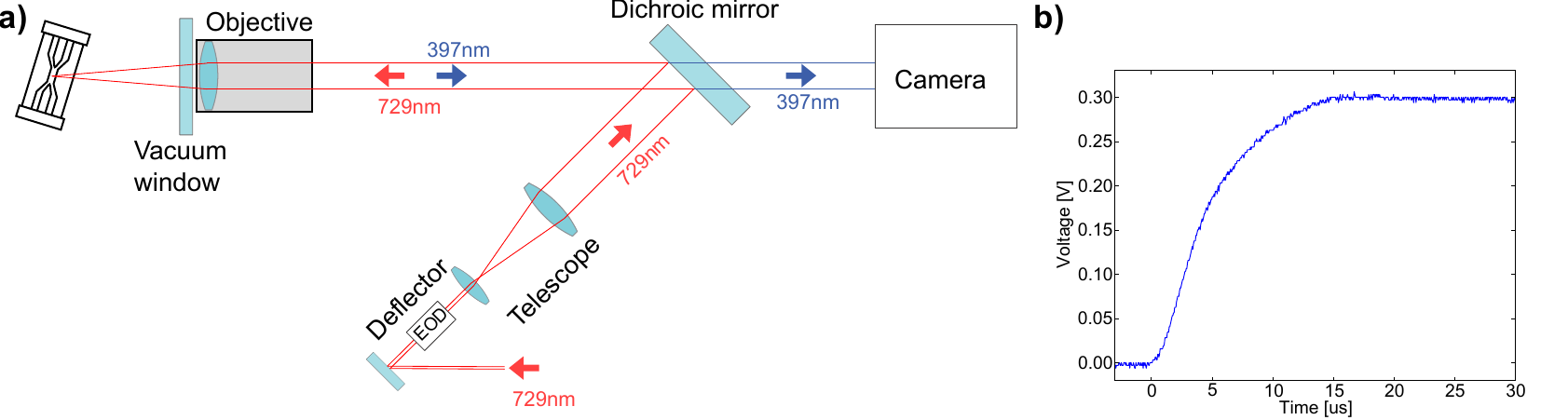}
  \centering
  \caption{a) Optical setup for the addressing beam setup.  b) Time
    dependence of the voltage on the EOD switching between two
    neighboring ions. After approximately $15 \mu s$ the voltage
    settles and thus the switching process is finished.}
  \label{fig:OpticAddress}
\end{figure}

The angle between the addressed beam and the trap axis is
$67.5^{\circ}$ which results in a smaller Lamb-Dicke parameter of
$\eta_{add}=2.5 \%$.  The addressed beam needs to be able to resolve
the individual ions in the string which means that the beam size needs
to be smaller than the inter-ion distance of approximately $5\mu
m$. This small beam size is realized with the aid of a custom high
numerical aperture objective situated in an inverted viewport as
sketched in figure~\ref{fig:OpticAddress}a).  Additionally, the beam
has to be rapidly switched between the ions which is realized with an
electro-optical deflector (EOD). The switching speed depends on the
capacitance of the EOD and the output impedance of the driving high
voltage amplifier. Figure~\ref{fig:OpticAddress}b) shows the voltage
on the EOD during a switching event between two neighboring ions which
demonstrated that a switching event requires approximately $15 \mu
s$. Experience has shown that a delay between the switching event and
the next light pulse of $30 \mu s$ is sufficient to switch between
arbitrary ions in a string of up to 8 ions. Note that the voltage ramp
measured at the EOD can only serve as an indicator for the position of
the laser beam but does not provide information about the settling
time of the laser light phase at the position of the ion. It was
observed that the phase of the light field keeps changing for more
than $100 \mu s$ after a switching event. However, this does not
affect the qubit operations for our set of operations as the AC-Stark
shift does not depend on the phase of the light field as described in
section~\ref{sec:tools-quant-inform}.

\subsection{Experiment control}

\begin{figure}[hh]
  \centering
  \includegraphics[width=8cm]{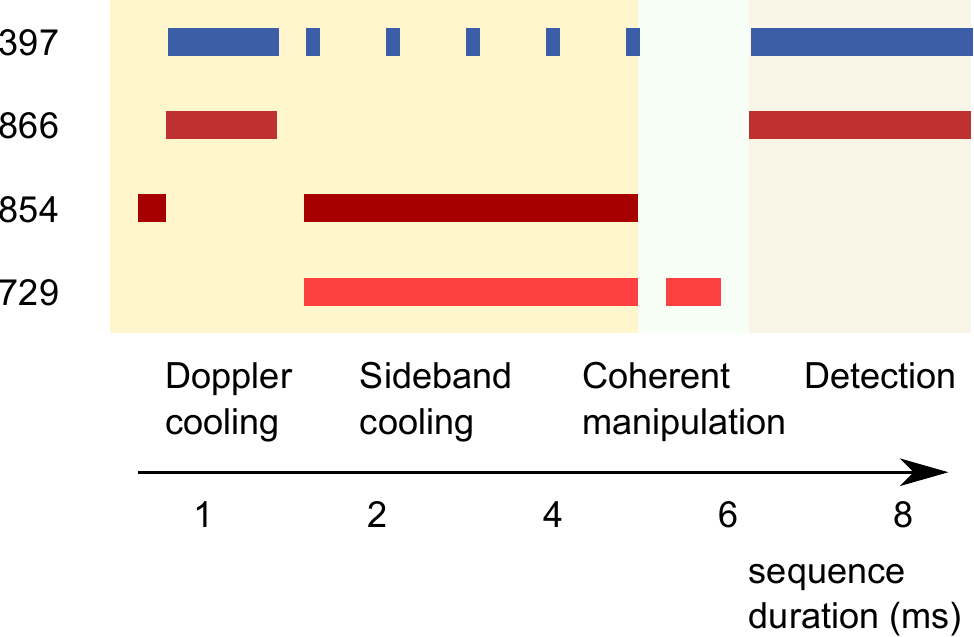}    
  \includegraphics[width=8cm]{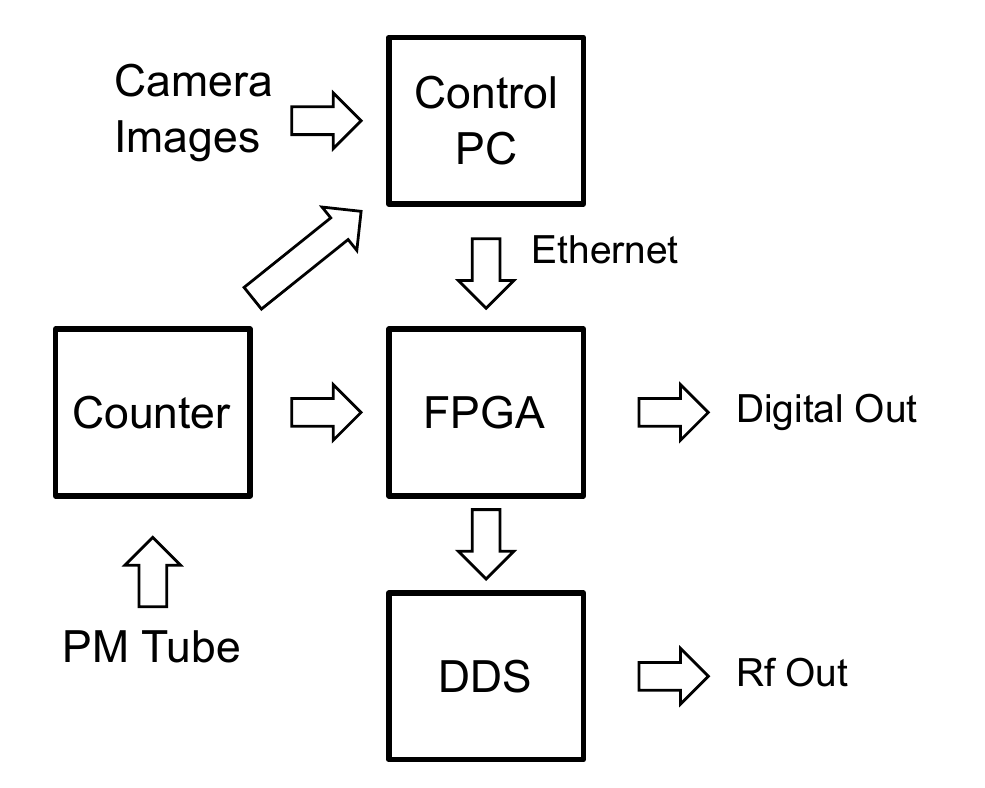}    
  \caption{a) Timing sequence of the different lasers for a typical
    experiment consisting of state initialization, coherent
    manipulation and measurement. b) Schematic representation of the
    experiment control hardware. The FPGA is programmed by the
    experimental control PC and controls the timing of all signals
    used in the experiment. RF signals for the coherent manipulation
    are generated by DDS. It is possible to perform conditional
    operations based on measurement outcomes with external counters
    that analyze the photon counts from the PMT.}
  \label{fig:ExpControl}
\end{figure}

Any ion-trap quantum information experiment requires precise and agile
control of duration, frequency and amplitude of laser beams
originating from multiple sources. A typical experimental sequence
consists of optical pumping, cooling the center-of-mass (COM) mode,
coherent operations and qubit measurements as shown in
figure~\ref{fig:ExpControl}a).  Usually the required control is
achieved by using acousto-optical devices which map the control over
intensity and frequency of the light field onto the manipulation of
amplitude and frequency of a radio-frequency signal. Thus, versatile
and fast radio-frequency(RF) signal generators are a necessity for a
high-fidelity quantum information processor. Modern RF signal
generators are commonly based on direct digital synthesizers (DDS)
enabling switching times on a nanosecond timescale and frequencies
between 1 and 400 MHz with sub-Hertz resolution. In our experiment,
these DDSs are controlled by a special purpose microcontroller
embedded on a field-programmable-gate-array
(FPGA)~\cite{Pham_2005}\footnote{http://pulse-programmer.org}. This
FPGA is able to generate digital pulses with a duration from 10ns up
to several seconds. In order to allow coherent rotation on different
transitions to be realized, the control system needs to be able to
perform phase-coherent switching between multiple frequencies. The
phase stability of the phase-coherent switching has been tested to be
$0.0001(90)^{\circ}$~\cite{Chwalla2009Absolute}. The controller is
connected to the experimental control computer via a standard ethernet
connection. For quantum algorithms requiring feed-forward operations,
such as teleportation, it is necessary to use the outcome of a
measurement within the algorithm to control subsequent operations in
the algorithm. This can be realized by analyzing the measurement
outcome by counting the PMT signal on dedicated counters and
performing the controlled operations in the sequence depending on
state of this counters~\cite{mark_tele}. A schematic view of the
control system including this feedback mechanism is shown in
figure~\ref{fig:ExpControl}b).

\begin{table}[hh]
  \centering
  \begin{tabular}{c|c|c|c}
    Parameter & Type & Required for each ion  \\ \hline
    Ion position & Voltage & yes \\
    Telescope lens position & Position & no \\
    Rabi frequency & Time & yes \\
    Zeeman splitting & Magnetic field & no \\ 
    Laser frequency drift  & Frequency & no \\
  \end{tabular}
  \caption{List of automatically calibrated parameters.}
  \label{tab:AutomatedParameters}
\end{table}

The FPGA determining the timing of the experiment is itself controlled
by a personal computer running a custom LabView program. This program
translates the sequence of operations from a human readable format to
binary code that can be executed on the FPGA. In order to minimize the
required time for calibrating the system, the parameters shown in
table~\ref{tab:AutomatedParameters} are calibrated automatically
without any user input.  Our set of operations can only be
implemented, if the frequency of the manipulation laser is close to
the qubit transition frequency. Since the frequency of each individual
transition is mainly determined by the center frequency of all
transitions shifted by the respective Zeeman shift due to the applied
magnetic field, it is sufficient to infer the magnitude of the
magnetic field and the frequency difference between the laser and the
center frequency. For this, the difference frequencies between the
laser and two distinct transitions are measured on the transitions
$4S_{1/2}(m_j=-1/2) \leftrightarrow 3D_{5/2}(m_f=-1/2)$ and
$4S_{1/2}(m_j=-1/2) \leftrightarrow 3D_{5/2}(m_f=-5/2)$ which allows
us to determine the long-term drift of the magnetic field and the
729~nm reference cavity. Typical values for the magnetic field drift
are $10^{-8} \, G/s$ and for the cavity drift 60~mHz/s which is
expected due to aging of the cavity spacer
crystal~\cite{Alnis2008Subhertz}.

In order to perform addressed single-qubit operations, the position of
the addressed beam with respect to the ion positions needs to be
characterized.  The position of the beam is controlled via the motorized
lens before the objective, as indicated in
figure~\ref{fig:OpticAddress}a), and the voltage that is applied to
the EOD. The calibration routine consists of moving the beam onto the
center of the ion string with the motorized lens, followed by finding
the EOD voltages for every individual ion. The position of the beam
with respect to the ions can be determined to approximately 50nm. In
order to perform the desired operations, the Rabi oscillation
frequencies on the global beam and the addressed beam need to be
measured. On the global beam, the Rabi frequencies of the two
transitions required for the drift compensation need to be calibrated,
whereas on the the addressed beam, the oscillation frequencies for
each ion for the AC-Stark operations are measured using Ramsey
spectroscopy. In general, the frequencies can be determined with a
precision of approximately 1\%.

\subsection{Measuring individual ions within a quantum register}
\begin{figure}[hh]
  \centering
  \includegraphics[width=8cm]{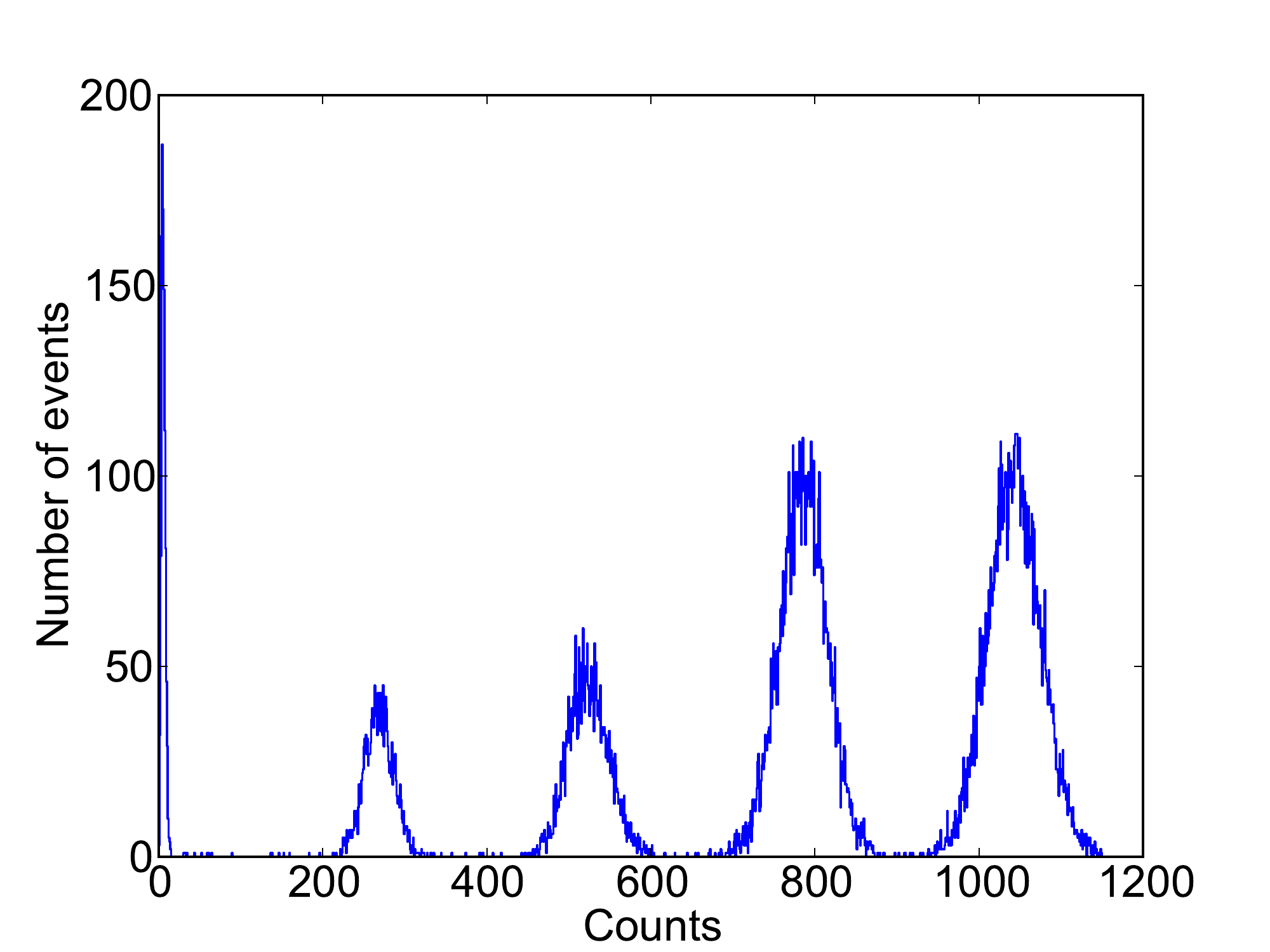}    
  \caption{Histogram of counted pulses from the PMT for a 4 ion
    string. The histogram is derived from 21200 measurements with a
    detection time of 5ms. }
  \label{fig:PMTHist}
\end{figure}

As described in section~\ref{sec:tools-quant-inform}, measuring the
quantum state of the ions is performed by counting single photons on
the $4S_{1/2} \leftrightarrow 4P_{1/2}$ transition.  We use high
numerical aperture objectives located in an inverted viewport to
reduce the distance between the ion and the objective as shown in
figure~\ref{fig:OpticTrap}. Two detection channels are available: one
with a photo-multiplier-tube (PMT) and another with an electron
multiplying CCD camera. The PMT integrates the photons over its
sensitive area and thus cannot infer any spatial information on the
ions. The number of detected photon counts depends on the number of
bright ions as is indicated in the histogram of PMT counts shown in
figure~\ref{fig:PMTHist}. By setting appropriate thresholds it is then
possible to determine the number of ions found in the
$4S_{1/2}=|0\rangle$ state which is sufficient information for
performing permutationally invariant state
tomography~\cite{Toth2010Permutationally} or determining the fidelity
of a multi-qubit GHZ state~\cite{GHZ}.

\begin{figure}[hh]
  \centering
  \includegraphics[width=16cm]{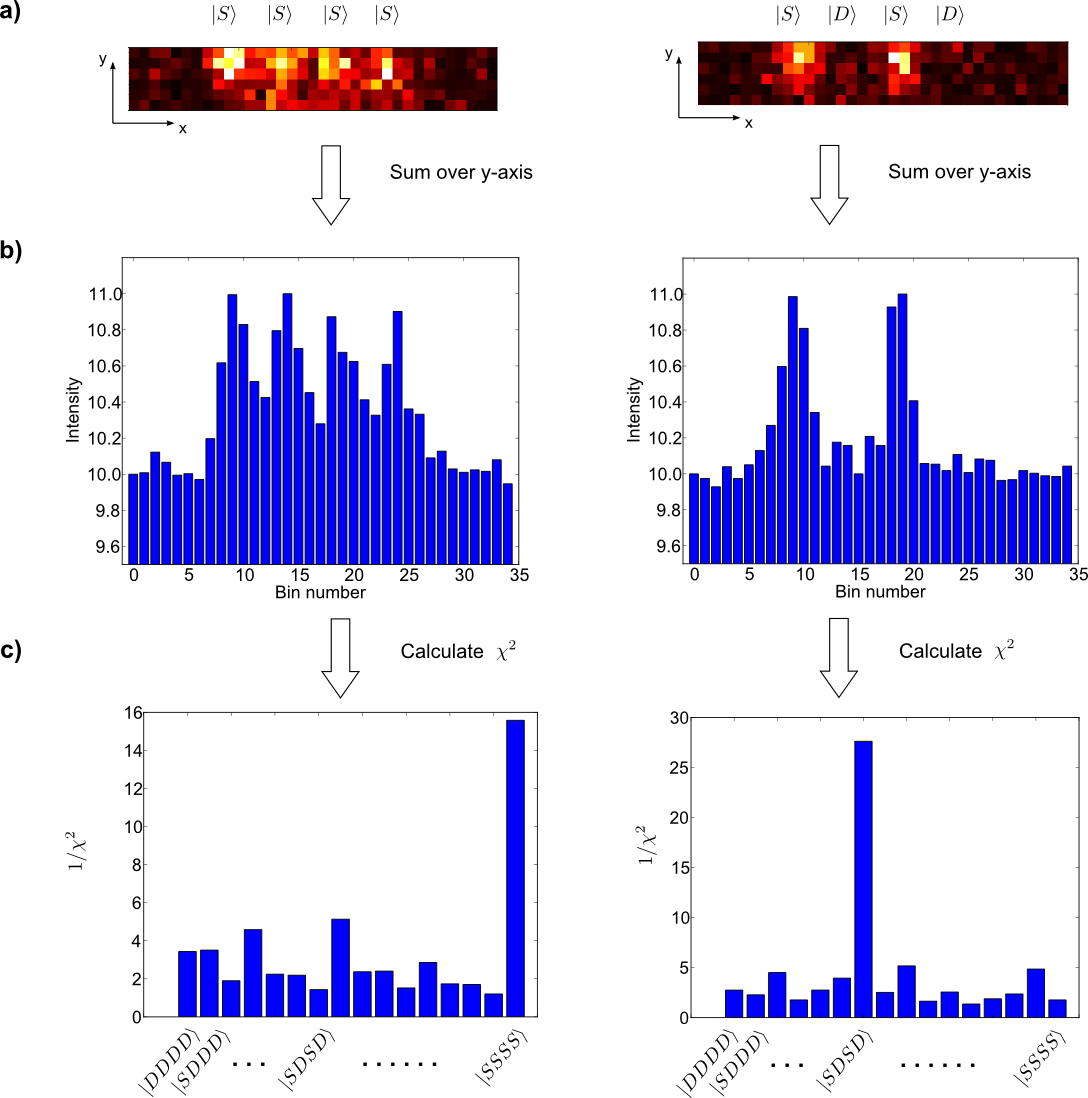}    
  \caption{Schematic illustration of the camera detection in a 4 ion
    register. (a) False color image of the region of interest. (b)
    Brightness information after summation over the y-axis of the
    image. (c) $1/\chi^2$ of the sum with generated data for every
    possible state. The peak corresponds to the most likely state. In
    this case index 6 (15), which corresponds to the state
    $|SDSD\rangle$ ( $|SSSS\rangle$), is the most likely state. }
  \label{fig:CameraDetection}
\end{figure}

In contrast, the CCD camera provides spatially resolved information of
the detected light and is thus able to determine the state of each ion
in the string separately. For fluorescence detection, the same
objective is used as for the focused 729nm beam is used. As sketched
in figure \ref{fig:OpticAddress}a), the light at 729nm and at 397nm
are separated by a dichroic mirror.  The analysis of the camera data
is performed in five steps: (i) A camera image is taken with an
exposure time of 7ms. The value of each pixel corresponds to the
number of detected photons.  (ii) For further analysis, a limited
region of interest (ROI) around the ion's position of the whole camera
image is used. For a register of 4 ions the ROI consists of 35x5
pixels but the ROI size needs to be adjusted to the length of the ion
string.  (iii) The pixel values are summed over the y-axis of the
ROI-image to get the brightness information along the ion string. (iv)
This brightness distribution is then compared to pre-calculated
distributions which are generated from a reference image where all
ions are bright. From this reference image, the position and
brightness distribution of each ion are determined.  The state of the
ion string is then inferred by comparing the summed pixel values with
the pre-calculated distributions of each possible outcome by
calculating the mean squared error $\chi^2$. Finally, (v) the state
with the smallest mean squared error is chosen to be the most likely
state.  Two examples of this analysis procedure are shown in
figure~\ref{fig:CameraDetection}. Note that this method is not
scalable as the number of pre-calculated distributions grows
exponentially with the number of ions. However recent work on state
detection in trapped ion system promises efficient detection
schemes~\cite{Burrell2010Scalable}.


\section{Error sources}
\label{sec:error-sources}

Any implementation of a quantum computer will be affected by errors
which ultimately need to be corrected with quantum error correction
techniques. Identifying and characterizing the noise sources are
therefore crucial steps towards a large-scale quantum information
processor. In this analysis we distinguish noise sources, that affect
a qubit used as a quantum memory, from additional error sources, that
occur when performing operations. For the presented error sources, we
describe the origin, present a method to characterize the magnitude,
and provide typical values for our experimental system.

\subsection{Errors in the qubit memory}

In general, errors affecting a qubit memory are described by a
combination of phase damping and amplitude
damping~\cite{nielsen_chuang}. In optical qubits, amplitude damping
corresponds to decay from the excited to the ground state whereas
phase damping destroys the phase of a superposition state but does not
alter the population of the qubit.  The lifetime of the excited qubit
is a fundamental property of the ion species and gives an upper limit
to the storage time of a quantum memory encoded in an optical
qubit. In the experiment, the lifetime of the excited state can be
reduced due to residual light fields depleting the $3D_{5/2}$ state
via another state, or by collisions with background gas
particles. This possible error source can be investigated by
confirming that the time constant of the exponential decay from the
$3D_{5/2}$ state is close to the natural lifetime of
1.168(7)s~\cite{Barton2000Measurement}. In our setup, we find a
lifetime of $\tau_1=1.13(5)s$~\cite{Nigg2013Experimental}.

The second noise type, phase damping, is usually investigated with
Ramsey spectroscopy which determines the coherence properties of a
superposition state~\cite{fsk-decoherence}.
There, the qubit is initially prepared in an equal superposition of
the two computational states by a $R_0(\pi/2)$ rotation. After a
certain storage time, a second rotation $R_\pi(\pi/2)$ is applied that
ideally maps the qubit back into the state $|1\rangle$. If the phase
$\phi$ of the second pulse $R_\phi(\pi/2)$ is varied with respect to
the first pulse, the probability of being in state $|1\rangle$ is a
periodic function of $\phi$. If the coherence of the state is decreased
due to phase damping, the second mapping pulse cannot reach the basis
states anymore which is observed as a decrease in the amplitude of the
oscillation. This loss of contrast corresponds directly to the
remaining phase coherence of the superposition which naturally
decreases with increasing storage time.

\begin{figure}
  \centering
  \includegraphics[width=8cm]{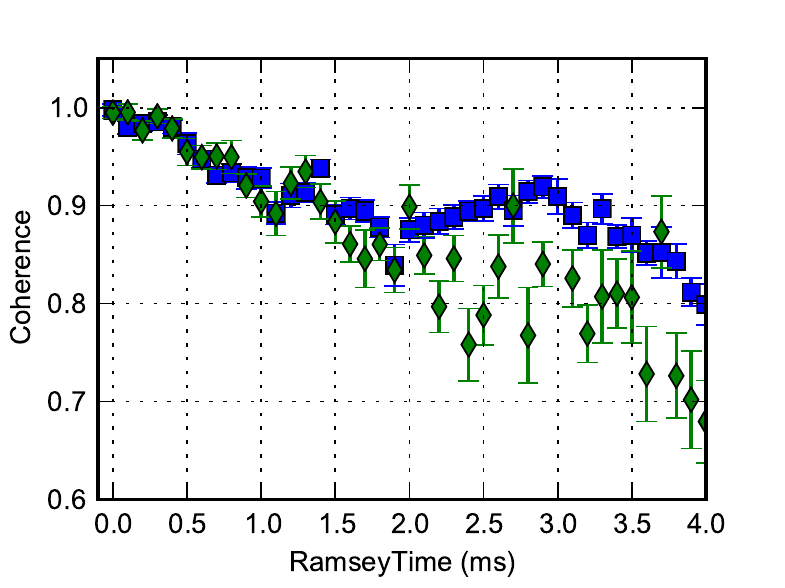}
  \includegraphics[width=8cm]{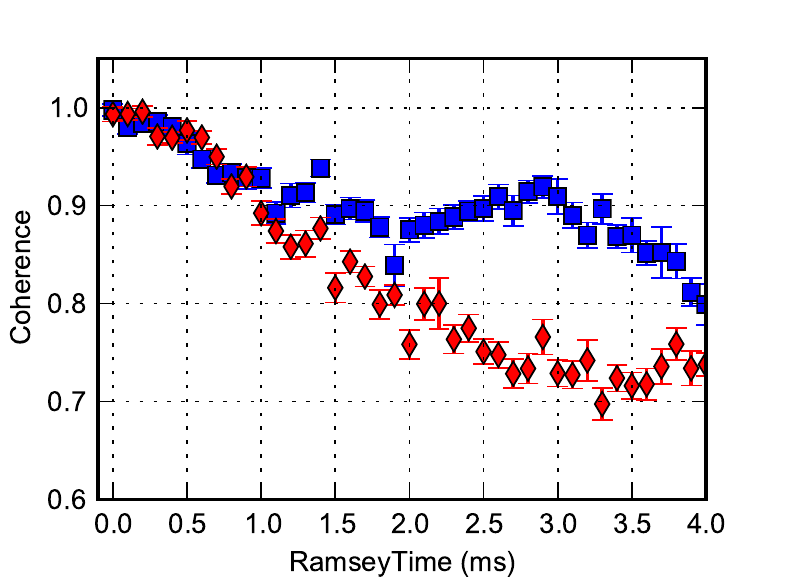}

  \caption{a) Ramsey contrast decay on two transitions with different
    sensitivity to the magnetic field fluctuations. Blue squares
    indicate the less sensitive $4S_{1/2} (m_j=-1/2) \leftrightarrow
    3D_{5/2} (m_j=-1/2)$ transition whereas green diamonds correspond
    to the $4S_{1/2} (m_j=-1/2) \leftrightarrow 3D_{5/2} (m_j=-5/2)$
    transition. b) Ramsey contrast decay on the transition which is
    least sensitive to magnetic field fluctuations, without (blue
    squares) and with (red diamonds) spin echo.}
  \label{fig:ramsey_echo}
\end{figure}

In our system, phase damping is predominantly caused by fluctuations
between the frequency of the qubit transition and the driving
field. The two main contributions are (i) laser frequency fluctuations
and (ii) fluctuations in the magnetic field which cause fluctuations
of the qubit transition frequency.  It is then possible to distinguish
between these contributions by investigating the coherence decay on
multiple transitions between different Zeeman substates of the
$4S_{1/2}$ and $3D_{5/2}$ levels because they show different
susceptibility to the magnetic field due to different Lande
g-factors. In figure~\ref{fig:ramsey_echo}a) the blue rectangles
represent the coherence decay on the $4S_{1/2} (m_j=-1/2)
\leftrightarrow 3D_{5/2} (m_j=-1/2)$ transition which is least
sensitive to fluctuations in the magnetic field. The green diamonds
show the coherence decay for the $4S_{1/2} (m_j=-1/2) \leftrightarrow
3D_{5/2} (m_j=-5/2)$ which has approximately 5 times higher
sensitivity to fluctuations of the magnetic
field~\cite{Chwalla2009Absolute,Tommaseo2003Mathsfg_scriptscriptstyle}. Note
that both transitions show effectively the same coherence decay for
storage times up to 1ms. This suggests that for typical experiments
where the coherent manipulation is shorter than 1ms, the main source
for dephasing are laser-frequency fluctuations.

The phase damping process can be theoretically described by a model
that applies random phase flips with a certain probability $p$ to
multiple copies of the same state. The ensemble of all states is then
described by a density matrix whose off-diagonal elements are affected
by the phase damping as $\rho_{i,j} \xrightarrow{i \neq j} \rho_{i,j}
(1-2p)$. This model of a phase-flip rate is close to the concept of a
bit-flip rate used in classical computer science and is therefore
widely used in theoretical works on quantum
information~\cite{nielsen_chuang}.  However, a physical model for
phase damping describes the phase-flip probability as a function of
the information storage time. In order to do so, one has to find a
noise-model describing temporal correlations of the noise source. The
most straightforward noise model assumes temporally uncorrelated noise
which leads to an exponential decay of the coherence characterized by
the transversal coherence time $\tau_2$ and therefore to off-diagonal
elements $\rho_{i, j} = \rho_{i,j} \,
e^{-t/\tau_2}$~\cite{Wineland1995Experimental}. This description is
used in most quantum computing models where the noise can be fully
characterized by the amplitude damping timescale $\tau_1$ and the
phase coherence time $\tau_2$~\cite{nielsen_chuang}.
In most physical systems, technical noise is temporally correlated and
thus this simple model of uncorrelated phase noise does not
apply~\cite{GHZ}.
In particular the coherence decay in our system deviates notably from
an exponential decay as can be seen in
figure~\ref{fig:ramsey_echo}a). This effect can be amplified with the
aid of a well known method to enhance the storage time of a quantum
memory known as the spin-echo technique. There, the basis states are
swapped at half the storage time which reverses the phase evolution
and thus cancels fluctuations provided their timescale is longer than
the storage time. However, it is possible that the performance with a
single echo is worse than the original register if this condition is
not satisfied. This effect is demonstrated in
figure~\ref{fig:ramsey_echo}b) where the coherence with spin-echo (red
diamonds) is worse than without echo (blue squares). There exist more
sophisticated methods to enhance the qubit storage time which are able
to take temporal correlations into account. A formal description of
this techniques is known as dynamical decoupling which has already
been demonstrated in various physical
systems\cite{Szwer2011Keeping,Sagi2010Process,Barthel2010Interlaced,deLange2010Universal,Ryan2010Robust,Du2009Preserving,Biercuk2009Optimized}.
For a given noise spectrum an optimal pattern of echo pulses can be
determined to maximize the phase coherence. Interestingly, one can use
this technique to determine the spectral noise density from multiple
coherence decays with varying number of
echos~\cite{Kotler2011Singleion,Bylander2011Noise}.
In the following we describe a simple experiment for identifying the
dominant features of the noise spectrum without using any spin echo
technique.

\begin{figure}
  \centering
  \includegraphics[width=12cm]{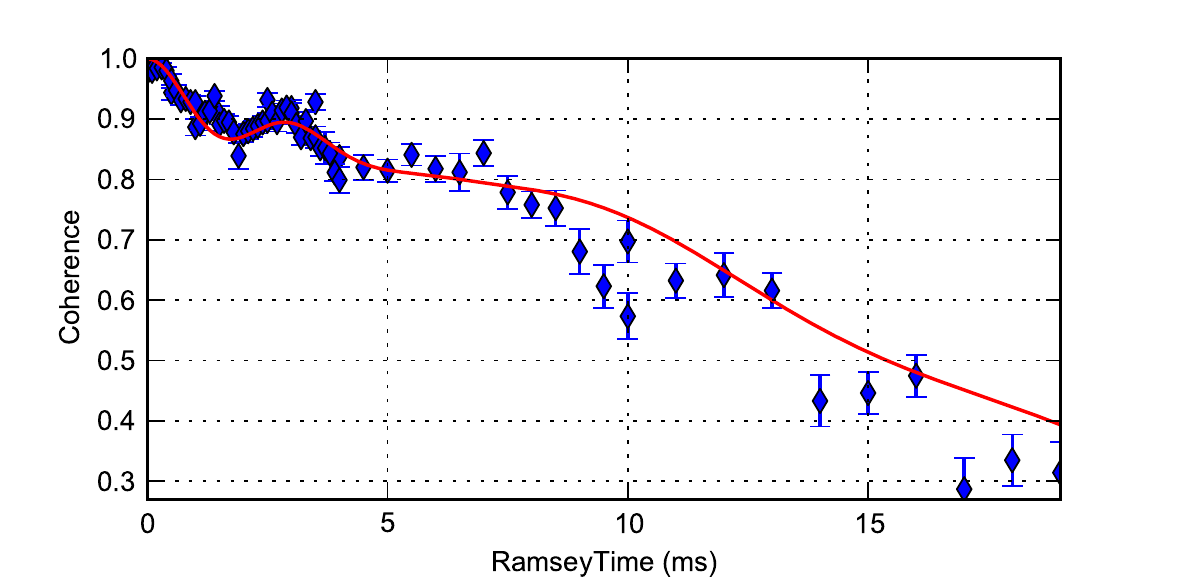}    
  \caption{Measured Ramsey contrast decay on the $4S_{1/2} (m_j=-1/2)
    \leftrightarrow 3D_{5/2} (m_j=-1/2)$ transition. The solid line
    shows a modeled Ramsey contrast decay with fitted parameters.}
  \label{fig:ramsey_decay_fit}
\end{figure}

It is possible to infer the noise spectrum from a coherence decay
$C(T)$ without any echo when only a few parameters of the noise
spectrum need to be determined.  For a given noise spectrum
$A(\omega)$, the Ramsey contrast decay is given by
$$C(T) = \exp{ \biggl \{ -\int_0^\infty d \omega \frac{A(\omega)^2}{\omega^2} \sin^2(\omega T/2)\biggr \} } \; .$$
which is a special case of the general coherence decay for dynamical
decoupling given in reference~\cite{Bylander2011Noise}.  Calculating
the noise spectrum from a measured coherence decay is not uniquely
possible, thus we characterize $A(\omega)$ assuming a certain spectral
shape of the noise and inferring only a few parameters. Our main
source of phase noise at relevant timescales smaller than 1ms seems to
be the laser frequency noise and thus we model the spectrum
accordingly.  Typically a laser spectrum is modeled as a Lorentzian
line, which we extend with two broad Gaussian peaks, where the first
originates from the laser locking electronics centered at 300Hz and
the second peak is attributed to the second harmonic of the power line
frequency at 100Hz. We model these two contributions with Gaussian
peaks $G_\nu(\omega) = \exp((\omega-\omega_0-\nu)^2/\sigma^2)$ where
$\sigma=10$Hz. The resulting spectral noise density for our model is
then
$$A(\omega) = \alpha \biggl( \frac{ \gamma^2}{\gamma^2+(\omega-\omega_0)^2} + a_1 G_{300}(\omega-\omega_0) + a_2 G_{100}(\omega-\omega_0) \biggr) \; .$$
Noise at the fundamental frequency of the power line (50Hz) is not
included in the model as it is not distinguishable from Gaussian noise
for storage times below 10ms.  Figure \ref{fig:ramsey_decay_fit} shows
the fitted coherence decay of the model with parameters $\alpha=89
\mathrm{\sqrt{Hz}}$, $\gamma=3 \; \mathrm{Hz}$, $a_1=0.22$ and
$a_2=0.02$.

\begin{figure}
    \centering
    \includegraphics[width=12cm]{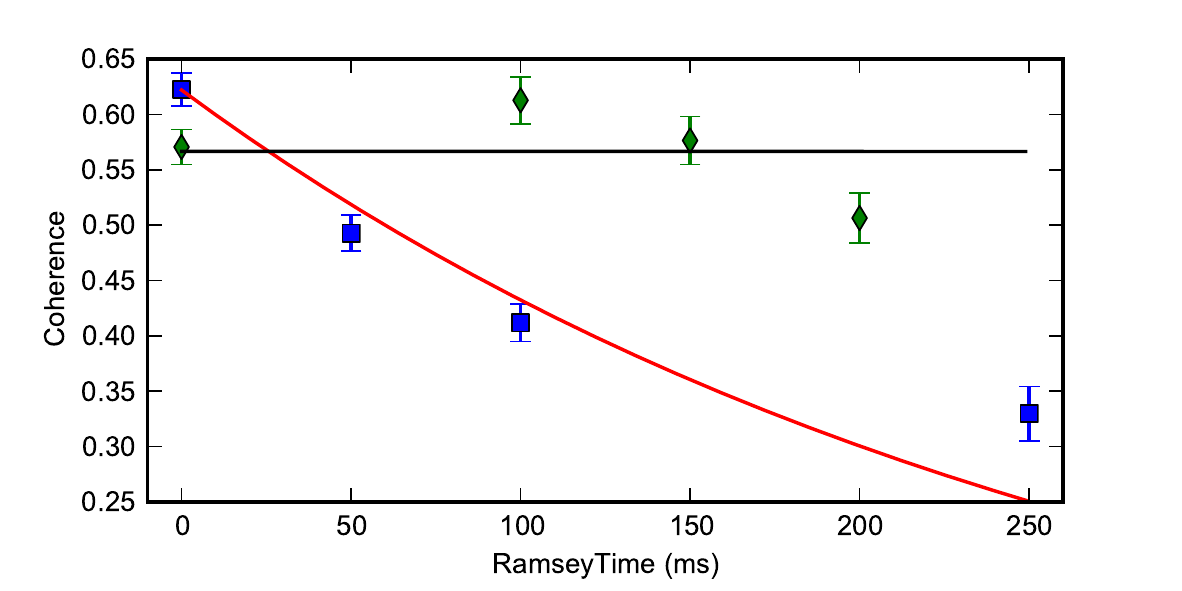}    
    \caption{Coherence as a function of the qubit storage time of an
      8-qubit DFS state encoded in the optical qubit (blue squares)
      and the ground-state qubit (green diamonds) eliminating
      amplitude damping decay. The solid lines represent the expected
      decay for both qubit types.}
    \label{fig:decay_8dfs}
  \end{figure}

When generalizing these results to multi-qubit systems, the spatial
correlation of the noise on all qubits needs to be considered. In our
system the noise from the laser and magnetic fields are almost
identical over the entire register and therefore the phase noise can
be modeled affecting the entire register simultaneously. This
correlation leads to a faster loss of coherence between states with
large total energy difference~\cite{GHZ}. On the other hand, this
spatial correlation enables decoherence free subspaces (DFS) which are
not affected by dephasing. The DFS consists of states in which
acquiring an equal phase on all qubits leads only to a global phase of
the state and thus to no dephasing. For example, a single logical
qubit can be encoded in two physical qubits as $|0_l\rangle =
|01\rangle + |10\rangle$ and $|1_l\rangle = |01\rangle - |10\rangle$
respectively. The two logical states have identical total energy
difference and thus form a DFS, where a universal set of operations
with two logical qubits has been demonstrated in our
system~\cite{Monz2009Realization}. However, it is not clear how well
the concept of a DFS can be extended to larger register sizes, and
thus we show the coherence decay of an 8-qubit DFS state of the form
$|00001111\rangle + e^{i \phi}|11110000\rangle$ in
figure~\ref{fig:decay_8dfs}. The state is generated by preparing the
qubit register in the state $|00001111\rangle$ and performing a
$MS_{\phi=0}(\pi/2)$ operation. If the DFS is also present for 8 ions,
the loss of coherence should correspond to the spontaneous decay of
the $3D_{5/2}$ state resulting in an exponential decay of the
coherence with timescale $\tau = \tau_1/n$ where $n=4$ is the number
of excited ions. This is illustrated in figure~\ref{fig:decay_8dfs}
showing the measured coherence decay and the expected decay, assuming
only spontaneous decay. Furthermore, the spontaneous decay can be
eliminated by encoding the qubit in the two substates of the
$4S_{1/2}$ level as introduced in
section~\ref{sec:tools-quant-inform}. The red squares in
figure~\ref{fig:decay_8dfs} show no noticeable decay during a storage
time of 200ms where limitations of the experiment control system (and
PhD students) prevent investigating longer storage times. The storage
time limit of this DFS is then given by fluctuations in the magnetic
field gradient and is expected to be in the 30s
regime~\cite{Haffner2005Robust}.

\subsection{Errors in quantum operations}

Performing operations on the qubit adds additional noise sources, and
thus the error rate of the entire algorithm cannot be described by
spontaneous decay and phase damping. We will now describe these
sources by their physical causes and categorize them by their
occurrence in (i) state initialization, (ii) coherent manipulation and
(iii) state detection.

\subsubsection{Initialization}

As described in section \ref{sec:tools-quant-inform} the qubit is
initialized by means of an optical pumping process towards the
$4S_{1/2}(m=-1/2)$ state using a circularly polarized laser beam
aligned parallel to the magnetic field. The possible error sources are
(i) imperfect polarization of the pumping light and (ii) misalignment
with respect to the magnetic field. The polarization quality is
determined by the quality of the polarization optics and the
birefringence caused by stress on the window attached to the vacuum
vessel. The quantization axis can be aligned by biasing the current in
the different magnetic field coils. The error probability of this
process can be measured by transferring the remaining population from
the $4S_{1/2}(m=1/2)$ to the $3D_{5/2}$ level and measuring it
subsequently. If the transfer works perfectly, the population left in
the $4S_{1/2}$ level is due to imperfect optical pumping. Since the
transfer is imperfect, the population needs to be shelved multiple
times to multiple substates in the $3D_{5/2}$ manifold. Every shelving
pulse is performed with an error rate of less than 1\% and thus the
error rate of two combined shelving pulses is on the order of
$10^{-4}$. With this technique, the fidelity of the optical pumping
process can be determined accurately.  We find a fidelity of the
optical pumping process of better than 99.1\%~\cite{Monz2011Quantum}.
The second optical pumping technique, as introduced in
section~\ref{sec:tools-quant-inform}, is frequency selective on the
qubit transition. Thus the direction of the magnetic field with
respect to the laser beam can be neglected which leads to a more
robust pumping. With this technique we find a pumping fidelity of
larger than 99\% ~\cite{Roos2006Designer}.

\begin{figure}
  \centering
  \includegraphics[width=12cm]{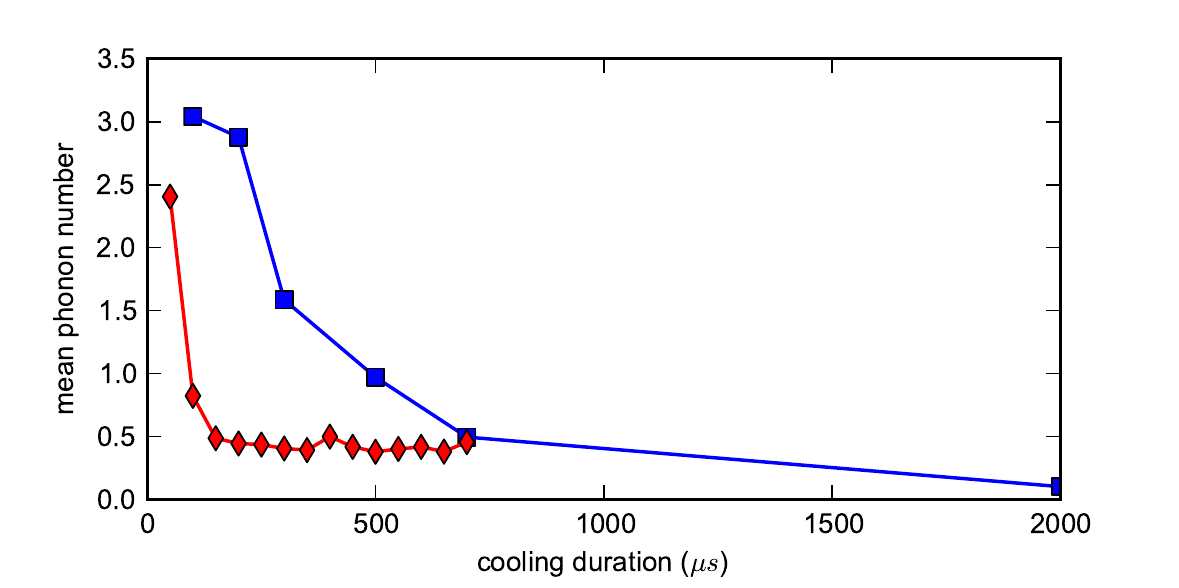}    
  \caption{Cooling rates for sideband cooling on the Raman (red
    diamonds) and the optical (blue squares) transition. Although
    cooling via the Raman process is faster it leads to a higher
    steady state phonon number.}
  \label{fig:raman_cool_results}
\end{figure}

The second initialization step prepares the ion in the motional ground
state of the harmonic oscillator. We treat the common-mode motion
(COM) separate from the other modes as it is used by the entangling MS
operations. In order to reach the lowest possible mean phonon number,
sideband cooling on the qubit transition as described in section
\ref{sec:tools-quant-inform} is performed on the common mode after a
Doppler pre-cooling cycle.  The final phonon occupation can be
determined by various techniques where a suitable method, when the
motion is close to the ground state, is to perform Rabi oscillations
on the motional sideband. This method uses the fact that the Rabi
frequency on the blue sideband for a given phonon number $n$ is given
by $\Omega_n = \sqrt{n+1} \, \eta \, \Omega_0$ where $\Omega_o$ is the
Rabi frequency on the carrier transition.  Rabi oscillations for a
given phonon distribution are described by
\[p_{|1\rangle} = \sum_n c_n \sin^2(\eta \Omega_0 / 2 \sqrt{n+1} \,
t)\] where the parameters $c_n$ can be determined by performing a
numerical fit to the measured data assuming a thermal distribution of
$ c_n={\langle n \rangle^n}/{(\langle n \rangle + 1)^{n+1}}$  which
is completely described by the mean phonon number $\langle n
\rangle$. A typical value for our experiments using sideband cooling
on the optical transition is $\langle n \rangle=0.05(3)$ after a
cooling time of 2~ms~\cite{Rohde2001Sympathetic}. In our setup we have
also the possibility of performing sideband cooling on the Raman
transition as introduced in section~\ref{sec:tools-quant-inform}.
This technique is used as an in-sequence recooling technique after a
measurement and therefore the cooling time has to be short compared to
the qubit coherence time.  Therefore, we adjust the cooling parameters
to achieve a faster cooling rate at the cost of a higher steady state
phonon number of $\langle n \rangle=0.5$ after a cooling time of $200
\mu s$. Figure~\ref{fig:raman_cool_results} compares the cooling rates
of the two distinct cooling techniques.

\begin{figure}
  \centering
  \includegraphics[width=8cm]{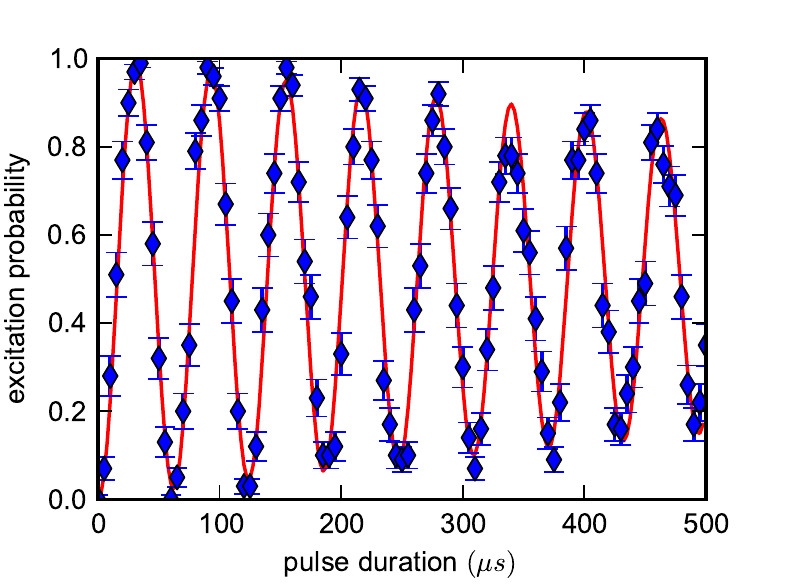}
  \includegraphics[width=8cm]{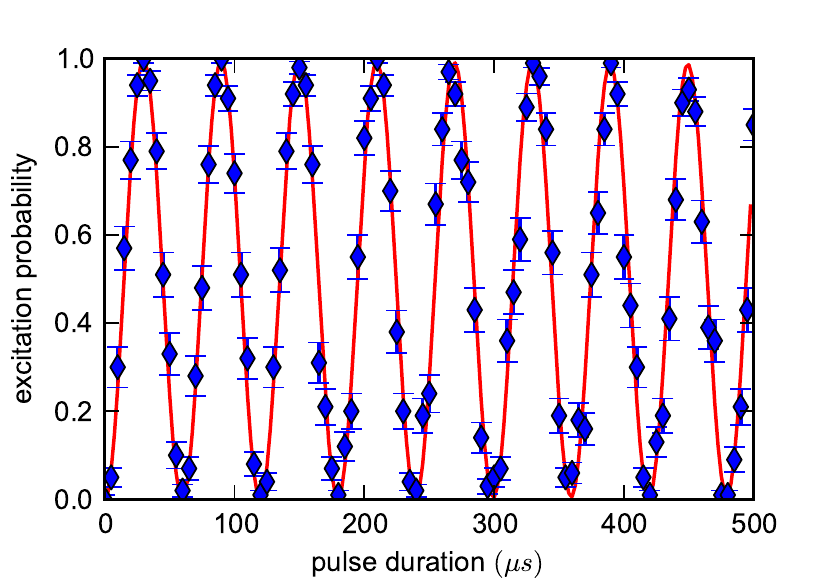}   
  \caption{Rabi oscillations in a three-qubit register illustrating
    the influence of thermal occupation of the motional modes when a)
    only the COM mode is cooled and b) all three axial modes are
    cooled.}
  \label{fig:rabi_cooling}
\end{figure}

In first-order Lamb-Dicke approximation ($\eta \ll 1$), the phonon
number of the remaining motional modes does not affect the dynamics of
the system. But as a second order effect, the occupation of these
modes alters the coupling strength of the ion to the light, which
causes an effective fluctuation of the Rabi frequency as the phonon
number follows a thermal distribution after
cooling~\cite{Wineland1995Experimental,Poschinger2012Interaction}. These
fluctuations are equivalent to intensity fluctuations of the driving
laser and cause a damping of the contrast of the Rabi oscillations.
This is illustrated in figure~\ref{fig:rabi_cooling}a) which shows
Rabi oscillations in a register of three ions where sideband cooling
was applied only to the COM mode. In contrast,
figure~\ref{fig:rabi_cooling}b) shows the same oscillations where all
three axial modes were cooled subsequently and the damping of the
oscillations is reduced. An $N$ ion crystal features $3N$ modes and
thus cooling all modes in a crystal gets increasingly difficult for
larger registers. Fortunately, cooling all modes of the crystal is not
always necessary because the mean-phonon number decreases with
increasing mode energy. Therefore we cool only the three modes
corresponding to the lowest energies to effectively suppress this
error source for up to 10 ions. In our setup this error source is
smaller on the addressed beam than the global beam, because the
Lamb-Dicke parameter is smaller as described in
section~\ref{sec:experimental-setup}.


\subsubsection{Coherent manipulation}

Additional errors occurring during the coherent manipulation of the
quantum information are mainly due to (i) laser intensity fluctuations
(ii) crosstalk and (iii) the limited coherence of the motional mode.

\begin{figure}
  \centering
  \includegraphics[width=8cm]{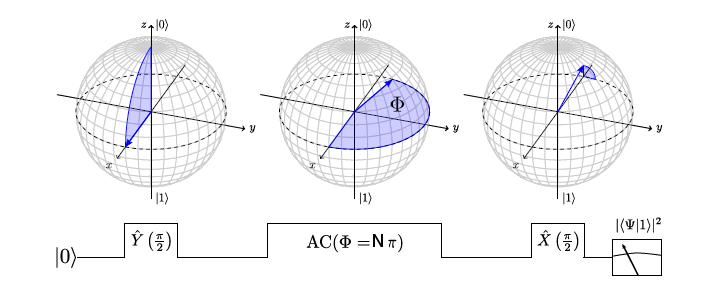}
  \includegraphics[width=8cm]{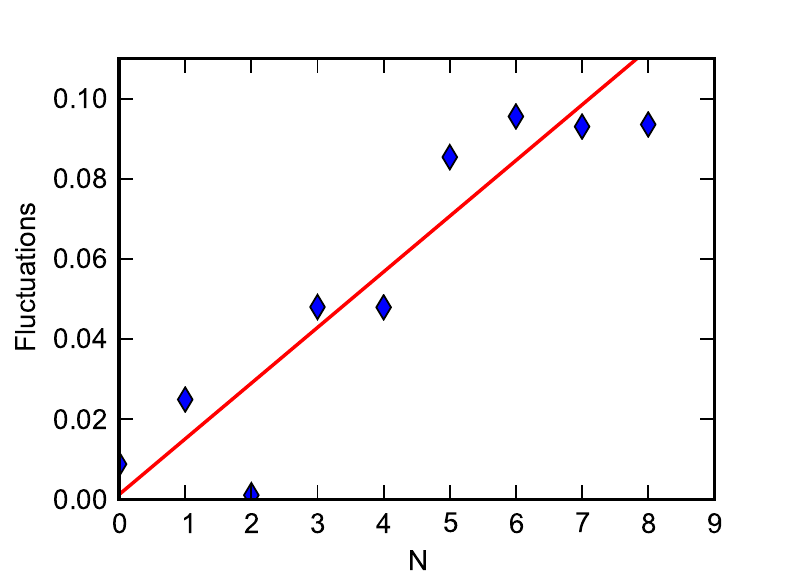}
  \caption{a) Measurement scheme for the slow intensity fluctuations
    with Ramsey type experiments. Multiple ($N$-times) rotations
    around the z-axis of the Bloch sphere are introduced into a Ramsey
    experiment translating intensity fluctuations into additional
    noise on the excitation probability. b) Measured state probability
    fluctuations $\Delta p$ for multiple $N$ where the slope is fitted
    to be $0.013(1)$ leading to effective intensity fluctuations of $
    \langle \Delta I/I \rangle _N = 0.41(6) \%$.}
  \label{fig:measure_int_fluct}
\end{figure}

Intensity fluctuations of the laser light manipulating the ions lead
to a fluctuating Rabi frequency and thus decrease the fidelity of the
operations.  Measuring the fluctuations of the light field with a
photodiode indicates that the fluctuations have relevant timescales
on the order of seconds to minutes. We assume therefore that the major
sources are (i) fluctuations of the coupling efficiency into a
single-mode optical fiber, (ii) thermal effects in acousto-optical
devices, (iii) polarization drifts in the fiber, which translate into
a varying intensity after polarization defining optics, and (iv) beam
pointing instability of the laser light with respect to the ion. These
intensity fluctuations can be measured directly on the ions by
inserting AC-Stark shift operations with varying length into a Ramsey
experiment as sketched in figure~\ref{fig:measure_int_fluct}a). The
AC-Stark shift operations convert intensity fluctuations directly
into phase fluctuations and thus the same Ramsey techniques as for
characterizing phase-noise can be used to measure them. The timescale
of the intensity fluctuations is slow compared to the required time
for taking 100 repetitions of the sequence and thus they cause excess
fluctuations of the measured excitation probabilities rather than a
coherence decay.

These excess fluctuations can be determined by comparing the standard
deviation of the measured data with the expected projection noise
$\Delta p^2 = \Delta_{proj}^2+ \Delta_{excess}^2$. This excess noise
in the state probability can be translated into fluctuations of the
rotation angle via error propagation.
We choose the rotation angle to be $\theta = N \pi$ with $N$ being an
integer yielding $\Delta \theta/\theta = \Delta p_{excess} / \pi N$
and perform this analysis up to $N=8$.  The measured state probability
fluctuations are then analyzed with a linear fit as shown in
figure~\ref{fig:measure_int_fluct}b).  From this, the relative
fluctuations of the rotation angles are determined which are directly
equivalent to the relative fluctuation of the Rabi frequency $\Delta
\theta/\theta=\Delta \Omega / \Omega$. For the AC-Stark shift
operations the Rabi frequency is directly proportional to the laser
intensity yielding $\Delta \Omega / \Omega= \Delta I/I$.  From the
fitted data we can identify the average laser fluctuations to be
$\langle \Delta I / I \rangle_N = 0.41(6)\%$.

\begin{figure}
  \centering
  \includegraphics[width=12cm]{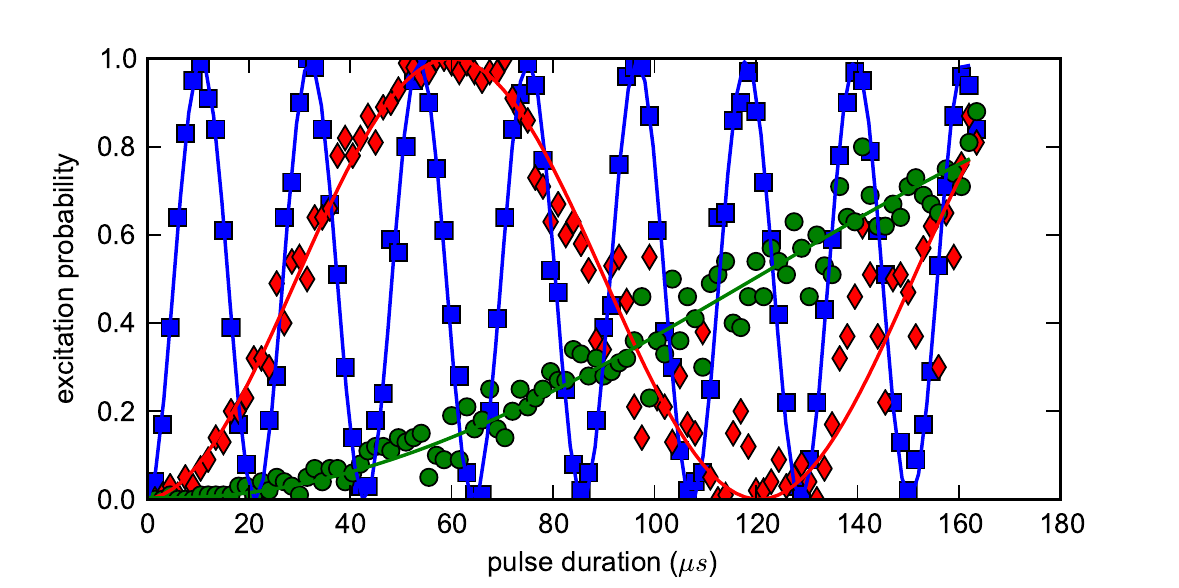}    
  \caption{Illustration of the crosstalk between neighboring qubits
    where the middle ion (blue rectangles) is addressed.  The fitted
    Rabi oscillation periods are 22$\mu s$ for the addressed ion 2,
    121$\mu s$ for ion 1 (red diamonds), and 464$\mu s$ for ion 3
    (green circles). }
  \label{fig:add_error}
\end{figure}

An error source that affects the register when performing addressed
single-qubit operations is crosstalk where due to the finite width of
the addressing laser, along with the desired ion, its neighboring ions
are affected. This addressing error is characterized by the ratio of
the flopping frequency of the addressed ion $i$ to the flopping
frequency of the ion $j$: $\epsilon_{i,j} = \Omega_i^2 /
\Omega_j^2$. The addressed operation, when addressing ion $i$, can
then be described by $S_z^{(i)}(\theta) = \exp(i \theta \sum_j
\sigma^{(j)}_z \epsilon_{i,j})$ where $\epsilon$ is the addressing
matrix describing the crosstalk. The magnitude of the error can then
be bounded by the maximum off-diagonal element of this matrix
$\epsilon_{max} = \max_{i \neq j} \epsilon_{i,j}$.  In
figure~\ref{fig:add_error} an example of excessive crosstalk in a
three ion register is shown with $\epsilon_{max}=22/121=18\%$. .
Typically, the maximum crosstalk on the addressed AC-Stark operations
is $\epsilon_{max} < 3\%$ for up to 8 ions where crosstalk between
more distant ions is typically smaller than $10^{-3}$.  Note that this
error is coherent, and thus can be undone if the whole addressing
matrix is known. Thus, the compensation of the crosstalk can be
integrated into the numerical optimization algorithm generating the
sequence of operations if the crosstalk is constant over time.

The presented error sources affect both, entangling as well as
non-entangling operations. A loss of coherence on the motional mode
does not affect non-entangling operations. However, the entangling MS
operation require coherences between different motional states, which
can be decreased by (i) fluctuations of the static voltages defining
the trap frequencies and (ii) heating of the ion string. The coherence
time of the motional mode can be determined by performing a Ramsey
type experiment which is only sensitive to the phase between two
different motional states. This is possible by using a superposition
of two motional states $1/\sqrt{2} (|S,0\rangle+|S,1\rangle)$ instead
of a superposition of the electronic state $1/\sqrt{2}
(|S\rangle+|D\rangle)$~\cite{Roos2008Nonlinear}. The motional
coherence is then measured analogous to the qubit storage time and yields
an exponential decay with time constant $\tau_{motion}=
110(20)\textrm{ms}$.
This coherence time is sufficiently long to allow high fidelity
operations to be performed~\cite{fsk-decoherence}.


\subsubsection{Measurement}

The dominant source of errors in the measurement of the qubit is given
by spontaneous decay from the $3D_{5/2}$ state during the measurement
process as well as stray-light. Both errors affect the measurement if
the qubit is projected into the $|0\rangle=3D_{5/2}(m=-1/2)$ state. The
stray-light is modeled by a Poissonian distribution with a mean value
of typically 1~counts/ms. The decay from the $3D_{5/2}$ state can also
be included which slightly modifies this distribution (for details
see reference~\cite{Roos2000diss}). For a ion being projected into the
$|1\rangle$ ($4S_{1/2}$) state, the photon distribution corresponds to
a simple Poissonian distribution with typically 50~counts/ms. The
detection error corresponds then to the overlap of the probability
distributions for a bright and a dark ion which can be well below
$10^{-3}$. The results from the CCD camera detection overlaps with the
PMT outcome at a level of better than 99.3\%~\cite{Riebe2005diss}.
\subsection{Estimating the effect of noise on an algorithm}

In order to determine the effect of the individual error sources for a
given sequence of operations, a numerical simulation including them
has to be performed on a classical computer, which is in general a
tedious task -- even for a few ions. We developed and use a numerical
modeling software named ``Trapped Ion Quantum Computing - Simulation
Program with Integrated Circuit Emphasis''
(TIQC-SPICE)~\cite{Wang2012Quantum}.  It follows a Monte Carlo
approach which simulates multiple random trajectories of varying
parameters~\cite{Dalibard1992Wavefunction,Molmer1993Monte} where each
trajectory yields a pure final state. The ensemble average over all
trajectories is then the density operator corresponding to the
simulated state of the system.

In the following we investigate two algorithms which show different
susceptibility to the individual noise sources. The algorithms will
not be explained in detail here as we focus on the effect of the
different noise sources on the fidelity of the final state.  As a
first algorithm we investigate a single timestep of an open-system
quantum simulator (details on the algorithm are given in
reference~\cite{Schindler2012Quantum}). This algorithm acts on two
system qubits but requires an additional auxiliary qubit whose state
can be neglected. We simulate its sequence of operations, shown in the
appendix, table~\ref{tab:SeqDDPT}, on a three-ion register using our
TIQC-SPICE program where each simulation run consists of a Monte-Carlo
simulation with 15 trajectories. The included noise sources and their
magnitudes are: crosstalk between next neighbors of
$\epsilon_{neighbor}=3\%$; Intensity fluctuations are given by $\Delta
I/I=2\%$; Dephasing is characterized by the coherence time
$\tau_{coh}=15 \textrm{ms}$ and the correlation time
$\tau_{corr}=333\mu s$ as defined in reference~\cite{GHZ}. Coupling to
spectator modes is modeled by additional intensity fluctuations of
$2\%$.  The simulated output state of the two system qubits is then
compared with the expected ideal state. The effect for each individual
noise source is identified by simulating the sequence multiple times
where for each simulation only a single source is affecting the
simulation.  The simulations for individual error sources indicate that
the dominant error source is dephasing as shown in
table~\ref{tab:TIQC}. This is expected because the duration of the
sequence of operations is 2~ms which is not short compared to the
coherence time of 15~ms.  Including all noise sources, the simulation
predicts a fidelity with the ideal density matrix of 79\% whereas the
experimentally measured fidelity is 72\%. The overlap of the simulated
with the measured density matrix is 94\%.

\begin{table}
  \centering
  \begin{tabular}{|c|c||c|}
    \hline
    Error source  &Overlap with ideal state \\ \hline
    All & 77 \% \\ \hline
    Crosstalk & 95\% \\ \hline
    Dephasing& 84 \% \\ \hline
    Intensity fluctuations & 99\% \\ \hline
    Spectator modes & 94\% \\ \hline
  \end{tabular}
  \caption{Results for the numerical simulation of a quantum 
    simulation algorithm where  smaller overlap means a larger error. In order to identify the dominant error source, 
    the simulation is performed multiple times with only a single active 
    error source. From the results one can infer that dephasing is 
    the dominant source of errors. The errors caused by motional heating,
    imperfect optical pumping and spontaneous decay are negligible.}
  \label{tab:TIQC}
\end{table}

The second simulated algorithm is a fully coherent quantum Fourier
transform (QFT) which is treated in more detail in section
\ref{sec:using-toolb-quant}.  The sequence of operation as shown in
the appendix, table~\ref{tab:SeqQFT}, is simulated with identical
parameters as the previous algorithm. The simulation predicts a
fidelity of 92.6\% with the ideal state whereas an experimentally
obtained density matrix leads to an overlap of 81(3)\%.  The
results of the simulation for the individual noise sources are shown
in table~\ref{tab:TIQC_QFT} where the biggest contribution is now
crosstalk.

\begin{table}
  \centering
  \begin{tabular}{|c|c|}
    \hline
    Error source & Overlap with ideal state \\ \hline
    All & 93 \% \\ \hline
    Crosstalk & 95\% \\ \hline
    Dephasing & 98 \% \\ \hline
    Intensity fluctuations & $>$99\% \\ \hline
    Spectator modes & $>$99\% \\ \hline
  \end{tabular}
  \caption{Results of a numerical simulation of a three-qubit QFT algorithm where
    a smaller overlap means a larger error. 
    Here, the dominant noise source is crosstalk. The errors caused by motional
    heating, imperfect optical pumping and spontaneous decay are negligible.}
  \label{tab:TIQC_QFT}
\end{table}

\section{Example algorithms}
\label{sec:using-toolb-quant}
\begin{figure}
  \centering
  \includegraphics[width=18cm]{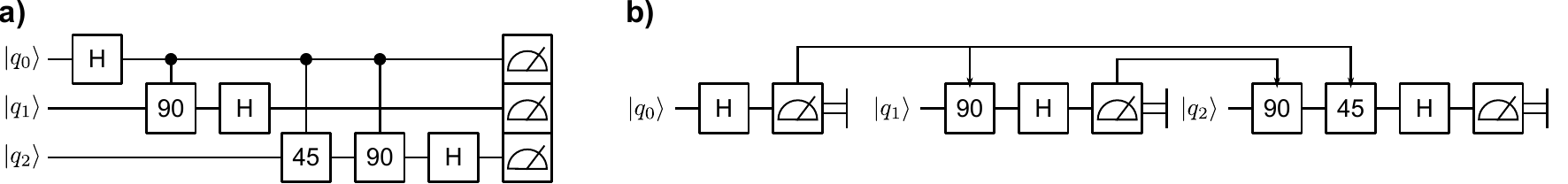}    
  \caption{(a) Quantum circuit for a three qubit QFT algorithm. (b)
    Single-qubit Kitaev version of the QFT. The measurement outcome is
    stored in a classical memory which controls the subsequent
    single-qubit rotations.}
  \label{fig:schem_qft}
\end{figure}

In the following we provide examples of how the available toolbox can
be employed to realize various quantum algorithms where we focus on
building blocks for a realization of Shor's algorithm to factor a
large integer numbers~\cite{Shor1994,nielsen_chuang}. The part of the
algorithm that requires a quantum computer is based on an
order-finding algorithm which itself requires the quantum Fourier
transform (QFT). This quantum analog of the discrete Fourier transform
maps a quantum state vector $|x\rangle= \sum_j x_j |j\rangle$, into
the state $|y\rangle= \sum_k y_k |k\rangle$ where the vector $y =
(y_1, \ldots ,y_N)=\mathcal{F}(x)$ is the classical discrete Fourier
transform of $x=(x_1 \ldots ,x_N)$~\cite{nielsen_chuang}. It is
straightforward to translate this operation into a quantum circuit
(see reference~\cite{nielsen_chuang}) where an example for three
qubits is shown in figure \ref{fig:schem_qft}a). The most
straightforward (although not necessarily the most effective) way to
implement the QFT is to realize directly the desired unitary using our
available operations. With our optimization toolbox as described in
section~\ref{sec:tools-quant-inform} we are able to find an optimized
decomposition of the three-qubit QFT consisting of 18 operations as
shown in the appendix, table~\ref{tab:SeqQFT}. The smallest MS
operation in the sequence is $\pi / 16$ and thus the MS operations has
to be optimized with this rotation angle. A maximally entangling
operation is then implemented by applying this operation 8 times
subsequently.

We benchmark the QFT by performing a full three-qubit quantum process
tomography and find a process fidelity of 72\% with the ideal
QFT~\cite{chuang_proctom}. However, in order to find the best suited
measure for the quality of an algorithm, one should consider how the
quantum algorithm is embedded in the given problem.  The QFT is almost
exclusively used as the final building block of larger algorithms and
then only the classical information of the final state is needed to
determine the algorithm's
performance~\cite{Griffiths1996Semiclassical}. The quantum process
fidelity is not the optimal measure to benchmark the performance of
the QFT as it includes correlations that do not affect the outcome of
the algorithm. One would rather choose a measure that utilizes the
classical probabilities of the individual output states which can be
described by a $2^N$ vector $p=(p_1, \ldots ,P_{2^N} )$.  Such a
measure is the squared statistical overlap (SSO) $S(p,q) = (\sum_i
\sqrt{p_i q_i})^2 $ which is the classical analog of the quantum state
fidelity~\cite{Fuchs1996Distinguishability}. An alternative suitable
measure for the classical information is the statistical
distinguishability $D(p,q)= 1 - 1/2 \sum_i |p_i - q_i|$, which is
related to the quantum trace distance. These benchmarks are applied to
a representative set of input states covering all possible periods. In
reference~\cite{Chiaverini2005Implementation}, a QFT algorithm was
benchmarked using five input states with different period and thus we use
similar input states for comparability, as shown in
table~\ref{tab:QFT}. The classical benchmarks yield on average an SSO
of 87\% which is considerably higher than the quantum process fidelity
of 72\%.

\begin{table}
  \centering
  \begin{tabular}{|c|c|c|c|}
    \hline
    Input state & Period& SSO & Distinguishability \\ \hline
    $1/\sqrt{8} \; (|111\rangle + |110\rangle + ... + |000\rangle)$ & 1& 77.1 & 77.1\\ \hline
    $1/\sqrt{4} \; (|110\rangle + |100\rangle + |010\rangle + |000\rangle)  $& 2 & 78.0 & 73.3\\\hline
    $1/\sqrt{4} \; (|110\rangle + |100\rangle + |011\rangle + |000\rangle)$ & 3 &90.4& 86.4\\\hline
    $1/\sqrt{2}  \;(|011\rangle + |000\rangle)$ & 4 &94.8 & 87.4\\\hline
    $|000\rangle$ & 8 & 97.3& 88.1 \\\hline
  \end{tabular}
  \caption{Results for a fully coherent 3 qubit QFT.}
  \label{tab:QFT}
\end{table}

Since the QFT is mainly used as the final block in an algorithm, it
can be replaced by the semi-classical QFT that exchanges the
quantum-controlled rotations by a measurement and a classically
controlled
rotation~\cite{Chiaverini2005Implementation,Griffiths1996Semiclassical}.
This requires the measurement of each qubit to be performed before the
operations that are controlled by this qubit. In
figure~\ref{fig:schem_qft} the time order of the measurements
corresponds to qubit $q_0$, $q_1$, $q_2$. A measurement furthermore
destroys all quantum coherence on the qubit and thus it is possible to
reuse the physical qubit and store the measurement outcome on a
classical computer.
This allows a semi-classical QFT to be performed on a single qubit as
sketched in figure~\ref{fig:schem_qft}b) which is known as the Kitaev
QFT~\cite{MartinLopez2012Experimental}.  Note that it is not possible
to generate an entangled input state with this version of the QFT and
thus the Kitaev QFT is more restricted than the semi-classical
QFT. Furthermore, the ability to measure and reset the qubit within
the algorithm is required, which is possible with our extended set of
operations.  In ion-trap systems, in-sequence measurements notably
disturb the motional state of the ion string and thus it is advisable
to make the measurement as short as possible. In this case we chose a
measurement duration of $150 \mu s$ which still allows for a detection
fidelity of $99 \%$~\cite{Schindler2013Undoing}. In order to achieve
high fidelity operations after such a measurement it appears necessary
to recool the COM mode with the Raman cooling technique as described
in section \ref{sec:tools-quant-inform}. In the special case of the
single-qubit QFT however only local operations are required after a
measurement which can furthermore be implemented with the addressed
beam. Due to the small Lamb-Dicke parameter, the quality of the
single-qubit operations is not notably affected by the thermal
occupation of the COM and the spectator modes after the measurements
and thus recooling is not required. In table~\ref{tab:QFTsemi} the
outcome for the single qubit QFT is shown for the non-entangled input
states used before, leading to an average SSO of 99.6\%.  As expected,
the single-qubit Kitaev QFT clearly performs better than the fully
coherent QFT.

\begin{table}
  \centering
  \begin{tabular}{|c|c|c|c|}
    \hline
    Input state & Period & SSO & Distinguishability \\ \hline
    $1/\sqrt{4}  \; (|000\rangle + |100\rangle + |010\rangle + |110\rangle$) & 2&99.5 & 94.5\\ \hline
    $1/\sqrt{2}  \;(|100\rangle + |000\rangle)$ &4 & 99.6 & 96.4\\ \hline
    $|000\rangle$&8 & 99.7 & 95.6\\ \hline
  \end{tabular}
  \caption{Results for the semiclassical Kitaev single qubit QFT.}
  \label{tab:QFTsemi}
\end{table}

\begin{figure}
  \centering
  \includegraphics[width=14cm]{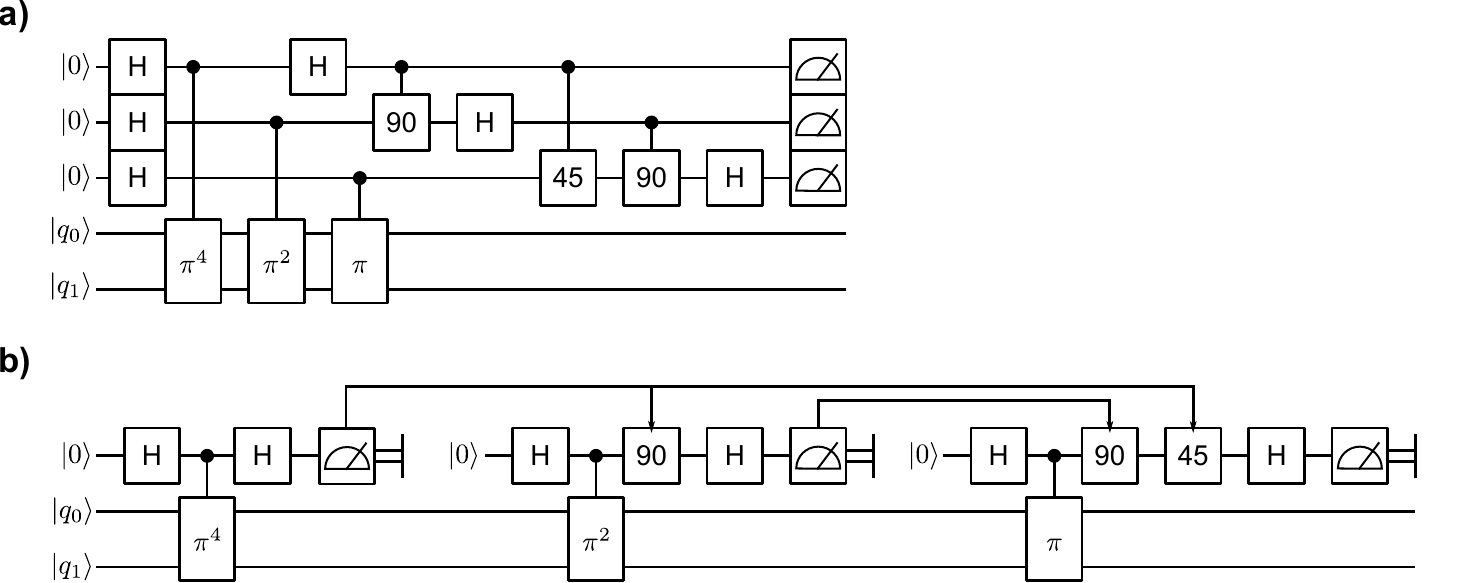}    
  \caption{order finding algorithm for a two qubit permutation
    operation in a) fully coherent and b) Kitaev version.}
  \label{fig:order_alg}
\end{figure}

One of the important algorithms that is compatible with the presented
Kitaev single-qubit QFT is the order-finding algorithm which is able
to determine the order of a permutation operation
efficiently~\cite{nielsen_chuang}. A permutation operation $\pi(y)$
has order $k$ if $k$-times application of the operations results in
the identity: $\pi(y)^k=y$, where $y$ is a decimal representation of
the $n$-qubit input state vector $|q_n \ldots q_0\rangle$. The
algorithm splits the available quantum register in two parts: (i) a
register where the permutation operation is applied and (ii) a QFT
register that is initially prepared in an equal superposition
state. The qubits from the QFT register control whether the
permutation operations are applied. This operation is analogous to a
CNOT operation where instead of the NOT operation the permutation
operation is controlled. The $k$-th qubit from the QFT register
controls the permutation operations $\pi(y)^l$ with $l=2^k$ as shown
in figure~\ref{fig:order_alg}a).  With this algorithm it is possible
to use the single-qubit QFT to reduce the number of required qubits
from 5 to 3 where the resulting quantum circuit is shown in
figure~\ref{fig:order_alg}b).

We seek to implement the optimized order-finding algorithm using
permutations on two qubits as a proof-of-concept experiment.  The
permutation operation is given by a unitary operation where we
implemented the operations shown in table~\ref{tab:order_unitary}
which span orders from 2 to 4. It becomes clear that the order of the
permutation can depend on the input state as, for example, $\pi_1(y)$
has order one for input states $y=|0\rangle,|2\rangle$ and order two
for $y=|1\rangle,|3\rangle$.  On the other hand, $\pi_2(y)$ shows
order two regardless of the input state. The complexity of the
algorithm depends on the investigated permutation operation, as the
controlled permutation operations require entangling operations. The
number of required operations for the individual permutation
operations are presented in table~\ref{tab:order_unitary} and the
sequences of operations can be found in the appendix. In contrast to
the single-ion QFT as presented above, the use of entangling
operations after measuring the QFT qubit is required. This makes it
necessary to recool the ion string within the sequence, where we
employ the Raman recooling technique as described in section
\ref{sec:tools-quant-inform}.  We choose a recooling time of $800 \mu
s$ as this proved to provide a good balance between remaining
excitation of the COM mode and additional phase damping due to the
cooling time~\cite{Schindler2013Undoing}.

\begin{figure}
    \centering
    \includegraphics[width=12cm]{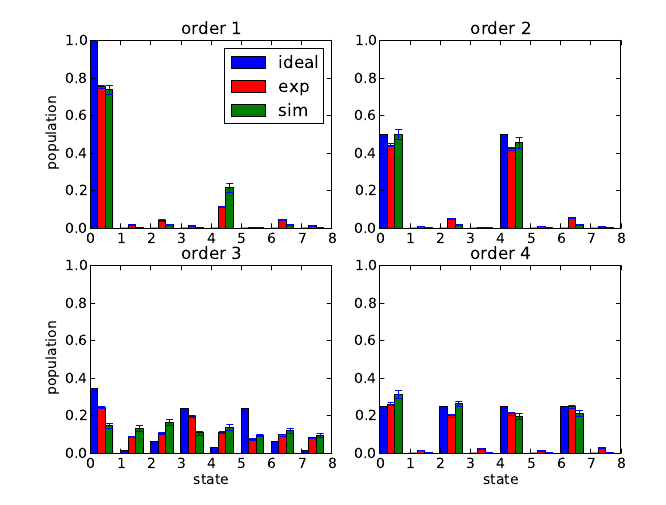}    
    \caption{State probabilities for the order finding algorithm for
      the permutation operations $\pi_1 \ldots \pi_4$. The ideal
      probabilities (blue), experimental results (red) and predictions
      from a classical simulation (green) are shown.}
    \label{fig:order_result}
  \end{figure}

The output of the algorithm is again classical and thus the classical
probabilities for measuring the state $|j\rangle$ are sufficient to
infer the quality of the operation. Figure~\ref{fig:order_result}
shows the classical probabilities of the basis states for all
permutation operations where the experimental results (blue bars) are
compared with the expected ideal probabilities (red bars) and
estimated probabilities from TIQC-SPICE simulations (green
bars). Again the implementation is benchmarked with the classical SSO
and distinguishability measures as presented in table~\ref{tab:Order}
yielding an average SSO of 80.7\%. The original problem is finding the
correct permutation and therefore one could think of using a classical
algorithm to find the most likely order for a given outcome. However,
finding such an efficient evaluation is beyond the scope of this work.

\begin{table}
  \centering
  \begin{tabular}{ccccc}
    y & $\pi_1(y)$ & $\pi_2(y)$ & $\pi_3(y)$ & $\pi_4(y)$ \\
    \hline
    $|0\rangle$ & $|0\rangle$ & $|1\rangle$ & $|0\rangle$ & $|3\rangle$ \\
    $|1\rangle$ & $|3\rangle$ & $|0\rangle$ & $|3\rangle$ & $|0\rangle$ \\
    $|2\rangle$ & $|2\rangle$ & $|3\rangle$ & $|1\rangle$ & $|1\rangle$ \\
    $|3\rangle$ & $|1\rangle$ & $|2\rangle$ & $|2\rangle$ & $|2\rangle$ \\
    \hline
    $\max (\mathrm{order})$ & 2 & 2 & 3 & 4 \\
    \hline
    no. of operations $\pi(y)$& 11 & 10 & 23 & 24 \\
    no. of operations $\pi(y)^2$& - & - & 17 & 10 \\

  \end{tabular}
  \caption{Representative unitary permutation operations for order 2 to 4 
    which were used as examples for the order-finding algorithm. The number
    of operations for applying the operation once and twice are also shown. The sequence of operation for the controlled permutation operations are presented in the appendix.}
  \label{tab:order_unitary}
\end{table}

\begin{table}
  \centering
  \begin{tabular}{|c|c|c|c|}
    \hline
    Order & Permutation operation & SSO & Distinguishability \\ \hline
    1 & $\pi_1(|0\rangle)$& 75.3(7) & 75.3(7) \\ \hline
    2 &$\pi_2(|0\rangle)$ &86.4(6) & 86.5(6)\\ \hline
    3 &$\pi_3(|1\rangle)$ &85.9(6) & 70.3(8)\\ \hline
    4 & $\pi_4(|0\rangle)$ &91.6(5) & 90.7(6)\\ \hline
  \end{tabular}
  \caption{Results for the semiclassical Kitaev order finding algorithm using the permutation operations defined in table~\ref{tab:order_unitary}.}
  \label{tab:Order}
\end{table}

\section{Conclusion and Outlook}

In conclusion we have presented a small-scale quantum information
processor based on trapped $^{40}$Ca$^+$ ions. A set of operations
beyond coherent operations, that is suitable for implementing
arbitrary Markovian processes, has been introduced. The major noise
sources of the system acting on a qubit memory and during operations
have been analyzed and their influence on different algorithms have
been discussed. It has been shown that the dominant source of errors
depends on the actual sequence of operations.  We used the entire set
of operations to realize an efficient implementation of the
order-finding algorithm.  Here, we have been able to reduce the number
of required qubits in the sense that only a single qubit is required
for the entire QFT register in the algorithm. We envision, that these
techniques will be the building blocks in a scalable implementation of
Shor's factoring algorithm. Furthermore we hope that the presented
methods for characterizing the noise sources will lead to successful
strategies for reducing the error rate in complex algorithms paving
the way to fault-tolerant quantum computation.

\section*{Acknowledgments}
We thank J. Home and M. M\"uller for helpful feedback on the
manuscript.  We gratefully acknowledge support by the Austrian Science
Fund (FWF), through the SFB FoQus (FWF Project No. F4002-N16), by the
European Commission (AQUTE), as well as the Institut f\"{u}r
Quantenoptik und Quanteninformation GmbH.  This research was funded by
the Office of the Director of National Intelligence (ODNI),
Intelligence Advanced Research Projects Activity (IARPA), through the
Army Research Office grant W911NF-10-1-0284. All statements of fact,
opinion or conclusions contained herein are those of the authors and
should not be construed as representing the official views or policies
of IARPA, the ODNI, or the U.S. Government.  \clearpage{}
\section*{References}
\bibliographystyle{bibgen}
\bibliography{longpaper}
\clearpage{}
\section{Appendix}

\begin{table}[h]
  \centering
  \begin{tabular}{|c|c||c|c|}
    \hline
    Number & Pulse & Number & Pulse\\
    \hline
1 & $S_z^{(1)}(1.5 \pi)$  & 11 & $MS_{1.5 \pi}(0.5  \pi)$  \\ 
2 & $R_\pi(1.5 \pi)$  &  12 & $S_z^{(3)}(1.75  \pi, 2)$  \\ 
3 & $MS_\pi(0.25)$  & 13 & $MS_{2.25 \pi}(0.5  \pi)$  \\ 
4 & $S_z^{(2)}( \pi)$  & 14 & $R_{1.75 \pi}(0.5  \pi)$  \\ 
5 & $MS_\pi(0.875  \pi)$  & 15 & $R_{2.25 \pi}(0.5 \pi)$  \\ 
6 & $S_z^{(3)}( \pi)$  & 16 & $MS_{2.25 \pi}(0.25  \pi)$  \\ 
7 & $MS_\pi(0.125  \pi)$  & 17 & $S_z^{(3)}(1.5  \pi)$  \\ 
8 & $S_z^{(3)}( \pi)$  & 18 & $S_z^{(2)}(1.0  \pi)$  \\ 
9 & $R_{0.5 \pi}(0.5  \pi)$  &19 & $R_{2.25 \pi}(0.5  \pi)$  \\ 
10 & $S_z^{(3)}(0.5  \pi)$  & &\\
    \hline
  \end{tabular}
  \caption{Sequence of operations for the algorithm used for an open system quantum simulator~\cite{Schindler2012Quantum}. }
  \label{tab:SeqDDPT}
\end{table}

\begin{table}[h]
  \centering
  \begin{tabular}{|c|c||c|c|}
    \hline
    Number & Pulse & Number & Pulse\\
    \hline
1 & $R_\pi(\pi/2)$ & 10 &  $R_{\pi/2}(3 \pi/16)$ \\
2 & $S_z^{(2)}(\pi)$ & 11 & $S_z^{(2)}(3\pi/2)$ \\
3 & $S_z^{(3)}(\pi /2)$ & 12 & $R_{\pi/2}(\pi/4)$ \\
4 & $MS_{\pi/2}(\pi/8)$ & 13 & $MS_{\pi/2}(\pi/8)$ \\
5 & $S_z^{(3)}(\pi)$ & 14 & $S_z^{(3)}(\pi)$ \\
6 & $MS_{\pi/2}(\pi/16)$ & 15 & $MS_{\pi/2}(\pi/8)$ \\
7 &  $R_{-\pi/2}(\pi/2)$ & 16 & $S_z^{(1)}(\pi/2)$ \\
8 & $S_z^{(2)}(\pi)$ & 17 & $S_z^{(2)}(\pi)$ \\
9 & $MS_{\pi/2}(3 \pi/16)$& 18 & $R_{\pi}(\pi/2)$ \\
    \hline
  \end{tabular}
  \caption{Sequence of operations for the fully coherent QFT operation on three qubits. }
  \label{tab:SeqQFT}
\end{table}

\begin{table}[h]
  \centering
  \begin{tabular}{|c|c||c|c|}
    \hline
    Number & Pulse & Number & Pulse\\
    \hline
    1 & $R_{\pi/2}(\pi/2)$  & 7 &  $MS_0(\pi/4)$ \\
    2 & $S_z^{(3)}(7 \pi /4)$ & 8 & $S_z^{(3)}(3 \pi /2)$ \\
    3 & $MS_0(\pi/2)$ & 9 &   $MS_0(\pi/2)$ \\
    4 & $R_{\pi}(\pi/2)$ & 10 & $R_{-\pi}(\pi/2)$ \\
    5 & $S_z^{(3)}(\pi /2)$ & 11 & $R_{-\pi/2}(\pi/2)$ \\
    6 & $R_{\pi}(\pi/4)$ & & \\
    \hline
  \end{tabular}
  \caption{Sequence of the controlled $\pi_1(y)$ permutation operation. }
  \label{tab:pi1}
\end{table}

\begin{table}[h]
  \centering
  \begin{tabular}{|c|c||c|c|}
    \hline
    Number & Pulse & Number & Pulse\\
    \hline
    1 & $R_{\pi}(\pi/2)$ & 5 & $MS_0(\pi/4)$ \\ 
    2 & $S_z^{(1)}(3 \pi /2)$ & 6 & $S_z^{(1)}(3 \pi /2)$\\
    3 & $MS_0(\pi/2)$ & 7 &  $R_{\pi}(\pi/2)$ \\
    4 & $R_{\pi}(\pi/2)$ & 8 & $S_z^{(2)}( \pi)$\\
    \hline
  \end{tabular}
  \caption{Sequence of the controlled $\pi_2(y)$ permutation operation. }
  \label{tab:pi1}
\end{table}

\begin{table}[h]
  \centering
  \begin{tabular}{|c|c||c|c|}
    \hline
    Number & Pulse & Number & Pulse\\
    \hline
    1 & $S_z^{(3)}(\pi /2)$ & 13 & $S_z^{(2)}(\pi/2)$ \\
    2 & $R_{\pi}(3 \pi/2)$ & 14 &  $S_z^{(3)}(3\pi/2)$ \\
    3 & $S_z^{(3)}(\pi /2)$& 15 & $MS_0(3\pi/4)$ \\
    4 &  $MS_0(\pi/4)$ & 16 & $R_{-\pi/2}(0.196 \pi)$ \\
    5 & $R_{-\pi}(5 \pi/2)$ & 17 &  $S_z^{(2)}(2\pi/3)$ \\
    6 & $S_z^{(1)}(3 \pi /2)$ & 18 & $R_{\pi/2}(0.196 \pi)$ \\
    7 & $R_{\pi}(\pi/2)$ & 19 & $R_{\pi}( \pi/4)$ \\
    8 & $R_{\pi/2}(\pi/4)$ & 20 & $MS_0(\pi/2)$ \\
    9 & $S_z^{(2)}(\pi)$ & 21 &   $S_z^{(2)}(7\pi/4)$ \\
    10 & $R_{\pi/2}(\pi/4)$ & 22& $R_{\pi/2}( \pi/2)$ \\
    11 &  $MS_0(\pi/2)$ & 23 &   $S_z^{(1)}(\pi/2)$ \\
    12 & $S_z^{(1)}(\pi)$ &&  \\
    \hline
  \end{tabular}
  \caption{Sequence of the controlled $\pi_3(y)$ permutation operation. }
  \label{tab:pi1}
\end{table}

\begin{table}[h]
  \centering
  \begin{tabular}{|c|c||c|c|}
    \hline
    Number & Pulse & Number & Pulse\\
    \hline
    1 & $R_{\pi/2}(\pi/2)$ & 10 & $S_z^{(2)}(3\pi /2)$ \\
    2 & $S_z^{(2)}(\pi /4)$ & 11 &  $MS_0(\pi/2)$ \\
    3 & $R_{-\pi}(\pi/2)$ & 12  & $S_z^{(3)}(3\pi /2)$ \\
    4 &  $MS_0(\pi/2)$ & 13 & $R_{\pi}(\pi/4)$ \\
    5 & $S_z^{(2)}(3\pi /2)$ & 14  &  $MS_0(\pi/4)$ \\
    6 & $MS_0(3\pi/4)$ & 15 & $S_z^{(1)}(3\pi /2)$ \\
    7 & $R_{\pi}(\pi/4)$ & 16 & $S_z^{(2)}(3\pi /2)$ \\
    8 & $S_z^{(3)}(\pi /4)$ & 17 & $R_{-\pi}(\pi/2)$ \\
    9 & $R_{\pi}(\pi/2)$ & & \\
    \hline
  \end{tabular}
  \caption{Sequence of the controlled $\pi^2_3(y)$ permutation operation. }
  \label{tab:pi32}
\end{table}

\begin{table}[h]
  \centering
  \begin{tabular}{|c|c||c|c|}
    \hline
    Number & Pulse & Number & Pulse\\
    \hline
    1 & $R_{-\pi}(\pi/2)$ & 13& $R_{-\pi/2}(0.196 \pi)$ \\
    2 & $R_{\pi/2}(\pi)$ & 14 & $S_z^{(1)}(4\pi /3)$ \\
    3 &  $S_z^{(1)}(3\pi /2)$ & 15 & $S_z^{(3)}(1.905\pi)$ \\
    4 &  $MS_0(7\pi/8)$ & 16 & $R_{\pi/2}(0.196 \pi)$ \\
    5 &  $S_z^{(3)}(\pi)$ & 17 & $R_{-\pi}( \pi/4 )$ \\
    6 & $MS_0(\pi/8)$ & 18 & $R_{-\pi/2}( \pi/2)$ \\
    7 & $R_{\pi/2}(\pi/2)$ & 19 & $MS_0(\pi/2)$ \\
    8 &$R_{-\pi}(3\pi/2)$ & 20 &  $S_z^{(2)}(\pi /3)$ \\
    9 &  $S_z^{(1)}(3\pi/2)$ & 21 &$MS_0(\pi/2)$ \\
    10 & $S_z^{(2)}(\pi/2)$ & 22 & $S_z^{(3)}(1.905\pi)$ \\
    11 & $MS_0(3\pi/4)$ & 23 &  $R_{\pi/2}( \pi/2)$ \\
    12 & $S_z^{(3)}(1.33 \pi)$ & 24  & $S_z^{(3)}(7 \pi/4)$ \\
    \hline
  \end{tabular}
  \caption{Sequence of the controlled $\pi_4(y)$ permutation operation. }
  \label{tab:pi4}
\end{table}

\begin{table}[h]
  \centering
  \begin{tabular}{|c|c||c|c|}
    \hline
    Number & Pulse & Number & Pulse\\
    \hline
    1 & $R_{\pi}(\pi/2)$ & 6 & $R_{\pi}(\pi/4)$ \\
    2 &  $S_z^{(1)}(3\pi /2)$ & 7 & $MS_0(\pi/4)$ \\
    3 &  $MS_0(\pi/4)$  & 8 & $S_z^{(1)}(3\pi/2)$ \\
    4 & $R_{\pi}(\pi/4)$ & 9 & $R_{\pi}(\pi/2)$ \\
    5   &  $S_z^{(2)}(\pi)$ & 10 & $S_z^{(2)}(\pi)$ \\

    1 & $R_{-\pi}(\pi/2)$ & 13& $R_{-\pi/2}(0.196 \pi)$ \\
    2 & $R_{\pi/2}(\pi)$ & 14 & $S_z^{(1)}(4\pi /3)$ \\
    3 &  $S_z^{(1)}(3\pi /2)$ & 15 & $S_z^{(3)}(1.905\pi)$ \\
    4 &  $MS_0(7\pi/8)$ & 16 & $R_{\pi/2}(0.196 \pi)$ \\
    5 &  $S_z^{(3)}(\pi)$ & 17 & $R_{-\pi}( \pi/4 )$ \\
    6 & $MS_0(\pi/8)$ & 18 & $R_{-\pi/2}( \pi/2)$ \\
    7 & $R_{\pi/2}(\pi/2)$ & 19 & $MS_0(\pi/2)$ \\
    8 &$R_{-\pi}(3\pi/2)$ & 20 &  $S_z^{(2)}(\pi /3)$ \\
    9 &  $S_z^{(1)}(3\pi/2)$ & 21 &$MS_0(\pi/2)$ \\
    10 & $S_z^{(2)}(\pi/2)$ & 22 & $S_z^{(3)}(1.905\pi)$ \\
    11 & $MS_0(3\pi/4)$ & 23 &  $R_{\pi/2}( \pi/2)$ \\
    12 & $S_z^{(3)}(1.33 \pi)$ & 24  & $S_z^{(3)}(7 \pi/4)$ \\
    \hline
  \end{tabular}
  \caption{Sequence of the controlled $\pi^2_4(y)$ permutation operation. }
  \label{tab:pi4}
\end{table}

\end{document}